\documentclass[pre,twocolumn,aps,groupedaddress]{revtex4-2}
\usepackage{graphicx}
\usepackage{bm}
\usepackage{amssymb,amsfonts,amsmath}%
\usepackage{color}

\newcommand{\bec}[1]{\mbox{\boldmath $ #1$}}
\newcommand{\meanN}{\overline{n}}
\newcommand{\meanS}{\overline{S}}
\newcommand{\meanrho}{\overline{\rho}}

\newcommand{\meanT}{\overline{T}}

\begin{document}
\title{Large-scale clustering of inertial particles in a rotating, stratified and inhomogeneous turbulence}
\author{Nathan Kleeorin$^{1,2}$}
\author{Igor Rogachevskii$^{1,3}$}
\email{gary@bgu.ac.il}
\homepage{http://www.bgu.ac.il/~gary}
\vspace{1mm}
\vspace{1mm}
\affiliation{
$^1$Department of Mechanical Engineering, Ben-Gurion University of
the Negev, P. O. Box 653, Beer-Sheva
84105, Israel \\
$^2$IZMIRAN, Troitsk, 108840 Moscow Region,  Russia\\
$^3$Nordita, KTH Royal Institute of Technology and Stockholm
University, Hannes Alfv\'ens v\"ag 12, SE-10691 Stockholm, Sweden}

\begin{abstract}
We develop a theory of various kinds of large-scale clustering of inertial particles in a rotating
density stratified or inhomogeneous turbulent fluid flows.
The large-scale particle clustering occurs in scales which are much larger
than the integral scale of turbulence, and it
is described in terms of the effective pumping velocity
in a turbulent flux of particles.
We show that for a fast rotating strongly anisotropic turbulence,
the large-scale clustering occurs in the plane perpendicular to rotation axis
in the direction of the fluid density stratification.
We apply the theory of the large-scale particle clustering
for explanation of the formation of planetesimals (progenitors of planets)
in accretion protoplanetary discs.
We determine the radial profiles of the radial and azimuthal components
of the effective pumping velocity of particles which
have two maxima corresponding to different regimes of the particle--fluid interactions:
at the small radius it is the Stokes regime, while at the larger radius it is the Epstein regime.
With the decrease the particle radius, the distance between the maxima increases.
This implies that smaller-size particles are concentrated nearby the central body of the accretion disk,
while larger-size particles are accumulated far from the central body.
The dynamic time of the particle clustering is about $\tau_{\rm dyn} \sim 10^5$--$10^6$ years,
while the turbulent diffusion time is about $10^7$ years,
that is much larger than the characteristic formation time of large-scale particle clusters ($\sim \tau_{\rm dyn}$).
\end{abstract}

\maketitle

\section{Introduction}

Turbulent transport of particles has been investigated
in a number of publications due to various applications
in geophysics, astrophysics, as well as in turbulent industrial flows
\cite{T77,C80,ZRS90,BLA97,SP06,ZA08,CST11,RI21}.
Particular interest is related to
transport of particles by rotating  turbulence
in astrophysical flows (e.g., formation of
planetesimals as progenitors of planets in protoplanetary disks
\cite{BR98,EKR98,PPK11,HUB16,HOP16,HOP16b}),
as well as in geophysical flows (e.g., dynamics of
particles in the atmospheric tornado and duststorms  \cite{L83}).

The key phenomena in particle turbulent transport include small-scale particle clustering and
large-scale particle clustering. Small-scale clustering arises in scales which are
much smaller than the integral turbulence scale. On the other hand,
the large-scale particle clustering  (i.e., formation of large-scale inhomogeneous spatial distributions
in particle number density) occurs in scales which are much larger than the integral scale of turbulence.

Typical examples of the large-scale particle clustering are
turbophoresis due to particle inertia and inhomogeneity of turbulence \cite{CTT75,RE83,G08,MHR18}
and turbulent thermal diffusion in temperature stratified turbulence \cite{EKR96,EKR97}.
Turbophoresis results in accumulation of inertial particles
in the vicinity of the minimum of the turbulent intensity, while
turbulent thermal diffusion causes the formation of large-scale particle
clusters in the vicinity of the mean temperature minimum.

Turbulent thermal diffusion has been intensively investigated analytically
\citep{EKR96,EKR97,EKR98b,EKRS00,EKRS01,PM02,RE05,AEKR17}
using different theoretical approaches.
This effect has been found in direct numerical simulations \citep{HKRB12,BRK12,RKB18},
detected in various laboratory
experiments \cite{BEE04, EEKR06,EKRL22,SKRL22,EKRL23},
studied in geophysical  \citep{SSEKR09}  and planetary \citep{EKPR97} turbulence as well as in
astrophysical turbulence \citep{HUB16}.

In the present  theoretical study, we show that the rotation results in various type of
large-scale clustering of inertial particles in density stratified or inhomogeneous
turbulent fluid flows.
For fast rotating strongly anisotropic turbulence, the dominant contribution to
the large-scale clustering occurs in the plane perpendicular to rotation axis
in the direction of the fluid density stratification.
The developed theory of large-scale particle clustering is applied
for explanation of formation of planetesimals (progenitors of planets) in accretion protoplanetary
discs.

This paper is organized as follows.
In Section II we discuss general ideas for developing a mean-field theory
describing turbulent transport of inertial particles in a rotating turbulence.
Here we present a general expression for effective pumping velocity
of particles in density stratified and inhomogeneous rotating turbulence.
In Section III we outline the method of derivations and approximations made
to determine the rotational contributions to
the Reynolds stress and the turbulent heat flux.
This allows us to derive expressions for various contributions to the effective pumping velocity
of particles describing different kinds of large-scale particle clustering
in small-scale density stratified and inhomogeneous rotating turbulence.
In Section IV we discuss applications of the analyzed effects
for formation of planetesimals in protoplanetary discs.
Finally, conclusions and discussions are given in Section V.
In Appendix A we derive expression for the total effective pumping velocity of inertial
particles, while in Appendix B we obtain the rotational contributions to the Reynolds stress and and the turbulent heat flux.
In Appendix~C we give the identities used for the derivation of the total effective pumping velocity of inertial
particles.

\section{Particles in a rotating turbulence}

Equation for the number density  $ n_{\rm p}(t,{\bm r}) $ of small particles advected
by a random fluid flow reads:
\begin{eqnarray}
{\partial n_{\rm p} \over \partial t} + {\rm div} \, (n_{\rm p} {\bm V})
= D \Delta n_{\rm p} ,
\label{CLA1}
\end{eqnarray}
where $ D $ is the coefficient of molecular (Brownian) diffusion,
${\bm V}$ is a random velocity field of
particles which they acquire in a turbulent fluid velocity ${\bm u}$
in a low-Mach-number flow.
Note that $ {\rm div} \,  {\bm V} \not= 0 $ is due to
particle inertia and inhomogeneity of the fluid density \cite{M87,EKR96,EKR98}.

Now we consider the large-scale dynamics of
inertial particles in a fast rotating turbulent fluid flow.
The equation for the mean field  $\meanN = \langle n_{\rm p} \rangle $
derived using different analytical approaches   \cite{EKR96,EKR97,RI21},
reads
\begin{eqnarray}
{\partial \meanN \over \partial t} + {\rm div} \,  \left[\meanN \, {\bm V}^{\rm eff} - \hat {\bm D}^{\rm T} \, {\bm \nabla} \meanN \right] = D \Delta \meanN ,
\label{CLA2}
\end{eqnarray}
where the first term in the squared brackets of
Eq.~(\ref{CLA2}), $\meanN \, {\bm V}^{\rm eff}$,
determines the contribution to the turbulent flux of particles caused by
the effective pumping velocity.
This effective pumping velocity is given by
\begin{eqnarray}
{\bm V}^{\rm eff} = - \left\langle \tau \,{\bm V} \,  {\rm div} \,  {\bm V} \right\rangle ,
\label{CLA3}
\end{eqnarray}
where $\tau$ is the turbulent correlation time of the fluid flow.
The second term, $- \hat {\bm D}^{\rm T} \,  {\bm \nabla} \meanN = - D_{ij}^{\rm T} \, \nabla_j \meanN$, in the squared brackets of
Eq.~(\ref{CLA2}) determines the contribution to the flux of particles caused by
turbulent diffusion, and
\begin{eqnarray}
D_{ij}^{\rm T} = \left\langle \tau \, V_i V_j  \right\rangle
\label{CLA4}
\end{eqnarray}
is the turbulent diffusion tensor.

We consider particles advected by a rotating turbulent
fluid flow for large P\'{e}clet numbers and large Reynolds numbers.
The equation of motion for particles is given by
\begin{eqnarray}
{d{\bm V} \over d t} = {{\bm u} - {\bm V} \over \tau_{\rm p}} + 2 {\bm
V} {\bm \times} {\bm \Omega} ,
\label{CLD1}
\end{eqnarray}
where ${\bm V}$ is the random velocity field of
particles caused by fluctuations of the fluid velocity ${\bm u}$, which are determined by equation
\begin{eqnarray}
{d{\bm u} \over dt} &=& - {{\bm  \nabla} P \over \overline{\rho}}  + \nu
\Delta {\bm u} + 2 {\bm u} {\bm \times} {\bm \Omega} + {{\bm f} \over \overline{\rho}} .
\label{CLD2}
\end{eqnarray}
Here ${\bm \Omega}$ is the angular velocity,
$P$ are fluid pressure fluctuations, $\overline{\rho}$ is the mean fluid of the density,
$\nu $ is the kinematic velocity, ${\bm f}$ is the external random force,
and $\tau_{\rm p}$ is the stopping time describing particle and fluid interaction.
The notion "stopping" time is originated in astrophysics.

When the mean-free path of the gas molecules is much smaller than the particle radius,
the particle and fluid interaction is described by the Stokes
regime with $\tau_{\rm p} = m_{\rm p} / (6 \pi \, \overline{\rho} \, \nu a_{\rm p})
= 2 \rho_{\rm p} \, a_{\rm p}^2/(9\overline{\rho}\, \nu)$, where $\rho_{\rm p}$
is the material density of particles,
$m_{\rm p}$ and  $a_{\rm p}$ are the particle mass
and the particle radius, respectively.
In the opposite case, when the mean-free path of the gas molecules is larger
than the particle radius, the stopping time $\tau_{\rm p}$ is determined by the Epstein regime
with $\tau_{\rm p} = \rho_{\rm p} \, a_{\rm p}/ (\overline{\rho} \, c_{\rm s})$ \cite{BR98,PPK11,HOP16,HOP16b},
where $c_{\rm s}$ is the sound speed.

Equation~(\ref{CLD1}) can be rewritten as
\begin{eqnarray}
A_{ij} \, V_j = u_i - \tau_{\rm p} \, {dV_i \over d t} ,
\label{CLD3}
\end{eqnarray}
where $A_{ij} = \delta_{ij} - \omega \, \varepsilon_{ijq} \hat \Omega_q$,
$\hat{\bm \Omega} ={\bm \Omega} / \Omega$ is the unit vector,
$\varepsilon_{ijk} $ is the fully anti-symmetric Levi-Civita unit tensor,
$\delta_{ij}$ is the symmetric Kronecker unit tensor
and $\omega = 2 \tau_{\rm p} \, \Omega$.
Multiplying Eq.~(\ref{CLD3}) by the inverse matrix
\begin{eqnarray}
A_{mi}^{-1} = {1 \over 1 + \omega^2}  \, \left(\delta_{mi} + B_{mi}\right) ,
\label{CLD0}
\end{eqnarray}
we obtain
\begin{eqnarray}
V_m = {1 \over 1 + \omega^2}  \, \left(\delta_{mi} + B_{mi}\right) \, \left[u_i - \tau_{\rm p} \, {dV_i \over d t}\right],
\label{CLD4}
\end{eqnarray}
where $B_{mi} = \omega (\varepsilon_{mis} \hat \Omega_s + \omega \, \hat \Omega_m \, \hat \Omega_i)$ and $A_{mi}^{-1} \, A_{in} = \delta_{mn}$.
By means of Eq.~(\ref{CLD4}), we determine ${\rm div} \, {\bm V}$ which characterises a compressibility
of particle velocity field  caused by the particle inertia, inhomogeneous fluid density and rotation:
\begin{eqnarray}
&& {\rm div} \, {\bm V} = {\rm div} \, {\bm u} + {1 \over 1 + \omega^2}  \, \biggl[\omega\,
\left(\hat{\bm \Omega}\cdot  {\rm rot} \, {\bm u}
- \omega \, {\rm div} \, {\bm u}_\perp\right)
\nonumber\\
&& \quad - {\rm div} \,  \left(\tau_{\rm p} \, {d{\bm V} \over d t} \right)  - \nabla_m \,  \left(\tau_{\rm p} \, B_{mi} \, {dV_i \over d t} \right) \biggr] ,
\label{CLD17}
\end{eqnarray}
where the fluid velocity ${\bm u}_\perp$ is in a plane perpendicular to ${\bm \Omega}$. To derive Eq.~(\ref{CLD17}), we use the identity $B_{mi} \, \nabla_m =\omega\, (\hat{\bm \Omega} {\bm \times} {\bm \nabla})_i + \omega^2 \,  \hat \Omega_i \, (\hat{\bm \Omega} \cdot {\bm \nabla})$.

There are three characteristic times in the system:
the stopping time describing particle - fluid interaction $\tau_{\rm p}$,
the correlation turbulent time $\tau_0$ in the integral scale $\ell_0$ and
the period of rotation $t_\Omega=2 \pi/\Omega$.
These three times are independent of each others.
In the developed theory there are two independent parameters: $\omega = 2 \tau_p \Omega$
and $\Omega_\ast = 4 \Omega \tau_0$.
The stopping time $\tau_{\rm p}$ is the smallest time in the system.
The case $\tau_{\rm p} \ll \tau_0 \ll \Omega^{-1}$ corresponds to a slow rotation $(\Omega\tau_0 \ll 1)$,
while the case $\tau_{\rm p} < \Omega^{-1}  \ll \tau_0$ corresponds to a fast rotation $(\Omega\tau_0 \gg 1)$.
For small time $\tau_{\rm p}$ (in comparison with $\tau_0$), we apply a method of iterations.

Using expression for the tensor $B_{mi}$ and applying a method of iterations
for small time $\tau_{\rm p}$, Eqs.~(\ref{CLD4})--(\ref{CLD17}) are transformed to
Eqs.~(\ref{CLD6}) and~(\ref{CLD20}) (see for details Appendix~\ref{appendix-A}).
Using Eqs.~(\ref{CLD6}) and~(\ref{CLD20}), we determine the total effective pumping velocity:
${\bm V}^{(\rm eff)} = - \left\langle \tau \,{\bm V} \,  {\rm div} \,  {\bm V} \right\rangle $,
which is given by
\begin{eqnarray}
&& {\bm V}^{(\rm eff)} = {\bm V}^{(\rm A)} + \hat {\bm \Omega} {\bm \times} {\bm V}^{(\rm B)}
+  \hat {\bm \Omega} \, \left(\hat {\bm \Omega} \cdot {\bm V}^{(\rm C)}\right) ,
\label{CLS0}
\end{eqnarray}
where
\begin{eqnarray}
&& {\bm V}^{(\rm A)} = {1 +3 \omega^2 \over (1 + \omega^2)^4} \, \biggl[(1 +3 \omega^2) \, {\bm V}^{(1)}
+ (1-\omega^2) \, \Big(\omega^2 \, {\bm V}^{(2)}
\nonumber\\
&& \; + {\bm V}^{(3)} \Big) - \omega^2 \,(3 +\omega^2) \, {\bm V}^{(4)}
+ 2 \omega^3 \, {\bm V}^{(5)} \biggr] ,
\label{EFV1}
\end{eqnarray}

\begin{eqnarray}
&& {\bm V}^{(\rm B)} = -{2 \omega^3 \over (1 + \omega^2)^4} \, \biggl[(1 +3 \omega^2) \, {\bm V}^{(1)}
- \omega^2 \, (1-\omega^2) \, {\bm V}^{(2)}
\nonumber\\
&& \;  + (1+ 5 \omega^2 + 2\omega^4) \, {\bm V}^{(3)}
- \omega^2 \, (3 +\omega^2) \, {\bm V}^{(4)} + 2 \omega^3 \, {\bm V}^{(5)} \biggr] ,
\nonumber\\
\label{EFV2}
\end{eqnarray}

\begin{eqnarray}
&& {\bm V}^{(\rm C)} = -{\omega^2 \, (1-\omega^2) \over (1 + \omega^2)^4} \, \biggl[
(1 +3 \omega^2) \, \Big({\bm V}^{(1)} + {\bm V}^{(3)} \Big)
\nonumber\\
&& \; - \omega^2 \, \Big({\bm V}^{(2)} + {\bm V}^{(4)} \Big)
+ 2 \omega^3 \, {\bm V}^{(5)} \biggr] ,
\label{EFV3}
\end{eqnarray}
and the effective velocities ${\bm V}^{(k)}$ describing various kinds of large-scale particle clustering,
are given by
\begin{eqnarray}
{\bm V}^{(1)} = - \left\langle \tau \,{\bm u} \,  {\rm div} \,  {\bm u} \right\rangle ,
\label{CLB1}
\end{eqnarray}

\begin{eqnarray}
{\bm V}^{(2)} = - \left\langle \tau \,{\bm u}  \left(\hat{\bm \Omega} \cdot {\bm \nabla}\right)
 \left(\hat{\bm \Omega} \cdot {\bm u}\right) \right\rangle ,
\label{CLB2}
\end{eqnarray}

\begin{eqnarray}
{\bm V}^{(3)} = - {\tau_{\rm p} \over \overline{\rho}} \, \left\langle \tau \,{\bm u} \,  \Delta P \right\rangle  ,
\label{CLB3}
\end{eqnarray}

\begin{eqnarray}
{\bm V}^{(4)} = - {\tau_{\rm p} \over \overline{\rho}} \, \left\langle \tau \,{\bm u} \,   \left(\hat{\bm \Omega} \cdot {\bm \nabla}\right)^2 P \right\rangle  ,
\label{CLB4}
\end{eqnarray}

\begin{eqnarray}
{\bm V}^{(5)} = - \left\langle \tau \,{\bm u} \,  \left(\hat{\bm \Omega} \cdot {\rm rot} \,  {\bm u}\right) \right\rangle  .
\label{CLB5}
\end{eqnarray}

\section{Large-scale effects}
\label{sect-3.1}

In this section, we derive the expressions for the effective velocities ${\bm V}^{(k)}$
defined by Eqs.~(\ref{CLB1})--(\ref{CLB5}).
We take into account an effect of rotation on turbulence.
In particular, to derive equation for the
rotational contributions to the Reynolds stress and the turbulent heat flux,
we follow the approach developed in \cite{RK18,RK19}.
We use equations for fluctuations of velocity ${\bm
u}'$ and entropy $s'= \theta/\overline{T} - (1-\gamma^{-1})P/\overline{P}$:
\begin{eqnarray}
{\partial {\bm u}' \over \partial t} &=& - \bec{\nabla} \biggl({P \over \overline{\rho}}\biggr) - {\bm g} \, s'
+ 2 {\bm u}' \times {\bm \Omega} + {\bm u}^{N},
\label{CKA1} \\
{\partial s' \over \partial t} &=& - ({\bm u}' \cdot \bec{\nabla}) \meanS + S^{N},
\label{CKA2}
\end{eqnarray}
where $P$ and $\theta$ are fluctuations of fluid
pressure and temperature, respectively,
$\gamma$ is the ratio of specific heats, and ${\bm g}$ is the acceleration due to the gravity.
The hydrostatic nearly isentropic basic reference state is defined by
$\bec{\nabla} \overline{P} = \overline{\rho} {\bm g}$, where
$\overline{T}$, $\overline{P}$, $\overline{S}$ and $\overline{\rho}$ are the mean fluid temperature, pressure, entropy,
and density, respectively, in the basic reference state.
In Eqs.~(\ref{CKA1})--(\ref{CKA2}),
${\bm u}^{N}=\langle ({\bm u}' \cdot \bec{\nabla}) {\bm u}'
\rangle - ({\bm u}' \cdot \bec{\nabla}) {\bm u}'$ and $S^{N}=\langle ({\bm u}' \cdot
\bec{\nabla}) s' \rangle - ({\bm u}' \cdot \bec{\nabla}) s'$
are the nonlinear terms,
and the angular brackets imply ensemble averaging.
In Eqs.~(\ref{CKA1})--(\ref{CKA2}) we neglect small
molecular viscosity and heat conductivity terms.
Equation~(\ref{CKA1}) is written in the reference
frame rotating with the constant angular velocity ${\bm \Omega}$.
The equations for fluctuations of
velocity and entropy are obtained by subtracting
equations for the mean fields from the
corresponding equations for the total velocity
and entropy fields.
The fluid velocity for a low Mach number
flows with strong inhomogeneity of the fluid density
$\overline{\rho}$ is assumed to be satisfied to the continuity equation written
in the anelastic approximation  ${\rm div} \, (\overline{\rho} \, {\bm u}') = 0$.

To study the effects of rotation on developed
density stratified inhomogeneous turbulence,
we perform the derivations which include the
following steps:
\begin{itemize}
\item{
using new variables for fluctuations of velocity ${\bm U} = \sqrt{\overline{\rho}}
\, {\bm u}' $ and entropy $s = \sqrt{\overline{\rho}} \, s'$;}
\item{
derivation of the equations for the
second-order moments of the velocity
fluctuations $\langle U_i \, U_j \rangle$ and the
entropy fluctuations $\langle s^2 \rangle$  in the ${\bm k}$ space;}
\item{
application of the multi-scale approach \cite{RS75}
that allows us to separate turbulent scales from large
scales;}
\item{
adopting the spectral $\tau$ approximation \cite{O70,PFL76,KRR90} (see Appendix~\ref{appendix-B});}
\item{
solution of the derived second-order moment equations in the
${\bm k}$ space;}
\item{
returning to the physical space to obtain expression for the Reynolds
stress and turbulent heat flux as the functions of the rotation rate $\Omega$.
The details of derivations are discussed in Appendix~\ref{appendix-B}.}
\end{itemize}

This procedure allows us to determine the rotational contributions
to the Reynolds stress and the turbulent heat flux,
and to derive expressions for the effective pumping velocities ${\bm V}^{(k)}$ of
inertial particles in rotating inhomogeneous or density stratified turbulence.
The latter describe various kinds of the large-scale particle clustering.

To introduce anisotropy of turbulent velocity field in the background turbulence caused by a fast rotation, we consider an anisotropic turbulence as a combination of a three-dimensional isotropic turbulence and two-dimensional turbulence in the plane perpendicular to the rotational axis. The degree of anisotropy $\varepsilon_u$ is defined as  the ratio of turbulent kinetic energies of two-dimensional to three-dimensional motions  [see Eq.~(\ref{CLF4}) in Appendix~\ref{appendix-B}].

The anisotropy parameter $\varepsilon_u$
appeared in the model of the background turbulence
depends on the Coriolis number Co=$2 \Omega \tau_0 = \Omega_\ast/2$,
where $\tau_0$ is the correlation time in the integral scale of turbulence.
For a slow rotation (small Coriolis numbers), the parameter $\varepsilon_u \to 0$.
For a fast rotation (large Coriolis numbers),
the parameter $\varepsilon_u \gg 1$.
In this case the background turbulence is a highly anisotropic nearly
two-dimensional turbulence, and the main rotational contributions
to the Reynolds stress are from the two-dimensional
part of turbulence.
Formally, in the present study when we consider a fast rotating turbulence,
the parameter $\varepsilon_u$ is not specified (it is a free parameter),
but it should satisfy the conditions $\varepsilon_u \gg 1$ for a fast rotation.
This phenomenological parameter can be determined, e.g., from DNS or
laboratory experiments.

\subsection{The effective pumping velocity ${\bm V}^{(1)}$ in inhomogeneous and density stratified turbulence}

We  determine the effective pumping velocity
${\bm V}^{(1)} = - \left\langle \tau \,{\bm u} \,  {\rm div} \,  {\bm u} \right\rangle$
that is given by Eq.~(\ref{CLB10}) in Appendix~\ref{appendix-A}.
For slow rotation ($\Omega_\ast^2 \ll 1$), the effective velocity ${\bm V}^{(1)}$ is given by
\begin{eqnarray}
&& {\bm V}^{(1)} = - {\bm \lambda} D_{T} +{1 \over 3} \, \Omega_\ast \biggl[ \hat{\bm \Omega} {\bm \times} \biggl({\bm \lambda}_\perp - {1 \over 2} {\bm \nabla}_\perp\biggr) \biggr]  D_{T} ,
\label{CLB11}
\end{eqnarray}
and for fast rotation ($\Omega_\ast^2 \gg 1$), it is given by
\begin{eqnarray}
&& {\bm V}^{(1)} = -  {3 D_{T} \over 4} \, {\bm \lambda}_\perp   .
\label{CLB12}
\end{eqnarray}
where  $D_T=\tau_0 \, \langle {\bm u}^2 \rangle / 3$ is the coefficient of turbulent diffusion,
${\bm \nabla}_\perp={\bm \nabla}-  \hat{\bm \Omega} \cdot {\bm \nabla}$,
${\bm \lambda} = - \meanrho^{\, -1} \, {\bm \nabla} \meanrho$ ,
${\bm \lambda}_\perp = {\bm \lambda} -  \hat{\bm \Omega} \cdot {\bm \lambda}$,
and $\Omega_\ast=4 \Omega \, \tau_0$.

It follows from Eqs.~(\ref{CLB11})--(\ref{CLB12}) that
in a density stratified turbulence there is a pumping effect of non-inertial particles
to the turbulent region with maximum mean fluid density.
This effect results in the accumulation
of non-inertial particles in the vicinity of the
maximum of the mean fluid density.
For a non-rotating turbulence, this phenomenon was studied in Ref.~ \cite{EKR97}.
The increase of the rotation rate increases anisotropy of turbulence,
and fast rotation causes the effective pumping velocity be directed
in the plane perpendicular to the rotation axis.

The physics of the accumulation
of non-inertial particles in the vicinity of the
maximum of the mean fluid density can be explained
as follows (see, e.g., book \cite{RI21}).
Let us assume that the mean fluid density
$\overline{\rho}_2$ at point $2$ is larger than the mean
fluid density $\overline{\rho}_1$ at point $1$. Consider
two small control volumes {``a''} and
{``b''} located between these two points,
and let the direction of the local turbulent
velocity in volume {``a''} at some instant
be the same as the direction of the mean fluid
density gradient $\bec\nabla \, \overline{\rho}$ (i.e.,
along the $x$--axis toward point $2$). Let the
local turbulent velocity in volume {``b''}
at this instant be directed opposite to the mean
fluid density gradient (i.e., toward point $1$).

In a fluid flow with a non-zero mean fluid density
gradient, one of the sources of particle number
density fluctuations, $n' \propto - \tau_0 \, \overline{n} \,
(\bec\nabla {\bf \cdot} \, {\bm u})$, is caused by a
non-zero $\bec\nabla\cdot{\bm u} \approx - {\bm u} \cdot
\bec\nabla \ln \overline{\rho} \not=0$.
Since fluctuations of the fluid velocity ${\bm u}$ are
positive in volume {``a''} and negative in
volume {``b''}, we have the negative divergence of the fluid velocity, $\bec\nabla {\bf
\cdot} \, {\bm u} < 0$, in volume {``a''}, and the positive divergence of the fluid velocity,
$\bec\nabla {\bf \cdot} \, {\bm u} > 0$, in volume
{``b''}.
Therefore, fluctuations of the
particle number density $n' \propto - \tau_0 \, \overline{n}
\, (\bec\nabla {\bf \cdot} \, {\bm u})$ are positive
in volume {``a''} and negative in volume
{``b''}. However, the flux of particles $n'\,
u_x$ is positive in volume {``a''} (i.e., it
is directed toward point $2$), and it is also
positive in volume {``b''} (because both
fluctuations of fluid velocity and number density
of particles are negative in volume
{``b''}). Therefore, the mean flux of
particles $\langle n' {\bm u} \rangle$ is directed, as is
the mean fluid density gradient $\bec\nabla \,
\overline{\rho}$, toward point~2. This results in formation large-scale
heterogeneous structures of non-inertial
particles in regions with a mean fluid density
maximum.

\subsection{The effective pumping velocity  ${\bm V}^{(2)}$}

We determine the effective pumping velocity ${\bm V}^{(2)} = - \langle \tau \,{\bm u}  (\hat{\bm \Omega} \cdot {\bm \nabla})
(\hat{\bm \Omega} \cdot {\bm u}) \rangle$ that is given by Eq.~(\ref{CLL1}) in Appendix~\ref{appendix-A}.
For slow rotation ($\Omega_\ast^2 \ll 1$), the effective velocity ${\bm V}^{(2)}$ is given by
\begin{eqnarray}
{\bm V}^{(2)} &=& {1 \over 4} \, \biggl[2{\bm \lambda}_\perp + \hat {\bm \Omega} \Big(\hat {\bm \Omega} \cdot{\bm \nabla} \Big)
+ {16 \over 15} \, \Omega_\ast \,\Big[\hat{\bm \Omega} {\bm \times}  \Big({\bm \lambda}_\perp
\nonumber\\
&&  - {1 \over 2} {\bm \nabla}_\perp\Big)\Big] \biggr] D_{T} ,
\label{CLL2}
\end{eqnarray}
and for fast rotation ($\Omega_\ast^2 \gg 1$), it is given by
\begin{eqnarray}
&& {\bm V}^{(2)} = - {3 \pi \over 8 \varepsilon_u} \, \biggl[\hat{\bm \Omega} {\bm \times}  \Big({\bm \lambda}_\perp - {1 \over 2} {\bm \nabla}_\perp\Big)\biggr] D_{T} ,
\label{CLL3}
\end{eqnarray}
where $\varepsilon_u$ is the degree of anisotropy of
turbulent velocity field which is large $(\varepsilon_u \gg 1)$ for fast rotation $\Omega_\ast \gg 1$, and it is small
for slow rotation ($\Omega_\ast^2 \ll 1$).
Since for fast rotation $\Omega_\ast \gg 1$, the degree of anisotropy $\varepsilon_u$ of
turbulent velocity field is large $(\varepsilon_u \gg 1)$, the effective pumping velocity ${\bm V}^{(2)}$
vanishes.

\subsection{The effective pumping velocity ${\bm V}^{(3)}$}

We determine the effective pumping velocity ${\bm V}^{(3)} = - (\tau_{\rm p} / \overline{\rho}) \, \left\langle \tau \,{\bm u} \,  \Delta P \right\rangle$ for inertial particles.
We consider a low-Mach-number flow, Ma $= u/c_{\rm s} \ll 1$,
where $c_{\rm s}$ is the sound speed. For a low-Mach-number
fluid flow, the turbulent fluid mass flux
$\langle \rho'  \, {\bm u} \rangle$ is very small, i.e., it is
of the order of ${\rm Ma}^2$, and fluctuations of the
fluid density $\rho'$ and velocity ${\bm u}$ are weakly correlated.
We use the equation of state for a perfect gas, which yields:
\begin{eqnarray}
{P \over \overline{P}} = {\rho' \over \overline{\rho}} + {\theta \over \overline{T}} + {\rm O}(\rho' \, \theta),
\label{CLG1}
\end{eqnarray}
and  $\langle P \, {\bm u} \rangle /\overline{P} = \langle \theta  \, {\bm u}
\rangle / \overline{T}$, where we neglect very small
turbulent fluid mass flux $\langle \rho'  \, {\bm u} \rangle$ for a low-Mach-number flow.
Here $\overline{P}$, $\,\overline{T}$ and $\overline{\rho}$ are the mean fluid pressure,
temperature and density, respectively.
Therefore, the effective pumping velocity for inertial particles is
\begin{eqnarray}
{\bm V}^{(3)} = - {\tau_{\rm p} \over \overline{\rho}} \, \left\langle \tau \,{\bm u} \,  \Delta P \right\rangle
\approx - {\tau_{\rm p} \,\overline{P}\over \overline{\rho} \, \overline{T}} \, \left\langle\tau {\bm u}({\bm x})\, \big[\bec{\nabla}^2 \theta({\bm x})\big] \right\rangle .
\nonumber\\
\label{CLG2}
\end{eqnarray}
Let us determine the correlation function $\left\langle \tau {\bm u}({\bm x})\, \big[\bec{\nabla}^2 \theta({\bm x})\big] \right\rangle$:
\begin{eqnarray}
&& \left\langle \tau {\bm u}({\bm x})\, \big[\bec{\nabla}^2 \theta({\bm x})\big] \right\rangle = \lim_{{\bm x}\to{\bm y}} \left\langle \tau {\bm u}({\bm x})\, \big[\bec{\nabla}^2 \theta({\bm y})\big] \right\rangle
\nonumber\\
&& \quad = - \int \tau(k) \, k^2 \,  \left\langle {\bm u}({\bm k}) \, \theta(-{\bm k}) \right\rangle^{(\Omega)} \, d{\bm k} ,
\label{CLG3}
\end{eqnarray}
where the correlation function $\langle u_i({\bm k}) \, \theta(-{\bm k}) \rangle^{(\Omega)}$ in the ${\bm k}$ space  follows from Eq.~(\ref{CLG6}).
Using Eqs.~(\ref{CLG6})--(\ref{CLLLG5}) given in Appendix~\ref{appendix-B} and Eqs.~(\ref{CLG2})--(\ref{CLG3}), we determine
the effective pumping velocity ${\bm V}^{(3)}$, that
is given by Eq.~(\ref{BCLG6}) in Appendix~\ref{appendix-A}.
For slow rotation, $\Omega_\ast^2 \ll 1$, the effective pumping velocity ${\bm V}^{(3)}$ is given by
\begin{eqnarray}
{\bm V}^{(3)}  &=&  - {2 \mu_0 \, D_T  \over 3} \, \ln{\rm Re} \, \biggl[{\bm \nabla}- \varepsilon_u \biggl({\bm \nabla} - {3 \over 4} \, {\bm \nabla}_\perp \biggr) \biggr]
\nonumber\\
&& \quad \times \ln \left(\overline{T} \, \overline{P}^{\, 1/\gamma-1}\right) ,
\label{CLG7}
\end{eqnarray}
and for fast rotation, $\Omega_\ast^2 \gg 1$, the effective pumping velocity ${\bm V}^{(3)}$ is given by
\begin{eqnarray}
{\bm V}^{(3)}  &=& - {\mu_0 \, D_T \over 2} \, \biggl[\ln{\rm Re} \, {\bm \nabla}_\perp
- \left({\pi \, \Omega_\ast \over \varepsilon_u}\right)  \, \hat {\bm \Omega} {\bm \times} {\bm \nabla}_\perp \biggr]
\nonumber\\
&& \quad \times \ln \left(\overline{T} \, \overline{P}^{\, 1/\gamma-1}\right) ,
\label{CLG8}
\end{eqnarray}
where $\mu_0 =  2 V_g \, L_P  / (3 D_T)$ with ${\bm V}_{\rm g} = \tau_p \, {\bm g}$
being the terminal fall velocity of particles and
$L_P = |{\bm \nabla} \overline{P} / \overline{P}|^{\, -1}$ being the height scale
of the mean pressure.
Here we use the equation of state for the ideal gas, $\overline{P} = R \, \overline{\rho} \, \overline{T}$,
with $R= k_{\rm B} /m_\mu$ being the gas constant, $m_\mu$ is the mass of a molecule
of the surrounding fluid and $k_{\rm B}$ is the Boltzmann constant.
We also take into account that in the basic reference state for the fluid
(i.e., in the hydrostatic equilibrium with a zero mean fluid velocity),
${\bm \nabla} \overline{P} \approx \overline{\rho} {\bm g}$
and  $\tau_p \, k_{\rm B} /m_\mu = V_{\rm g} \, L_P / \overline{T}$.
The latter expression follows from the identity for
the Stokes time for particles $\tau_p = \overline{\rho} \, V_{\rm g} \, L_P / \overline{P}$.

The effective pumping velocity ${\bm V}^{(3)}$ is related to the phenomenon
for turbulent thermal diffusion for inertial particles.
The mechanism of this effect is as follows.
The inertia causes particles inside the turbulent eddies to drift
out to the boundary regions between eddies due to the centrifugal inertial force.
Indeed, for large P\'eclet numbers, when molecular diffusion of
particles can be neglected in the equation for particle number density, it
follows that
\begin{eqnarray}
\bec\nabla {\bf \cdot} \, {\bm V} \approx - n_{\rm p}^{-1} \, \left[{\partial n_{\rm p} \over \partial t}
+ ({\bm V} \cdot {\bm \nabla}) n_{\rm p} \right] \equiv  - n_{\rm p}^{-1} \, {{\rm d}n_{\rm p} \over {\rm d}t} .
\nonumber\\
\label{ZZ30}
\end{eqnarray}
On the other hand, for inertial particles, $\bec\nabla {\bf \cdot} \, {\bm V}
= (\tau_{\rm p} / \rho)  \,\bec\nabla^2 P$. Therefore, in
regions with maximum fluid pressure fluctuations (where
$\bec\nabla^2 P < 0)$, there is an accumulation of
inertial particles, i.e.,  ${\rm d} n' / {\rm d}t
\propto - \overline{n} \, (\tau_{\rm p} /\overline{\rho}) \,\bec\nabla^2 P
> 0$. These regions have low vorticity and high
strain rate.
Similarly, there is an outflow of inertial
particles from regions with minimum fluid
pressure.

In homogeneous and isotropic turbulence with a zero gradient
of the mean temperature, there is no preferential direction,
so that there is no large-scale effect of particle accumulation,
and the pressure (temperature) of
the surrounding fluid is not correlated with the turbulent velocity field. The only
non-zero correlation is $\langle({\bm u} \cdot {\bm \nabla})P\rangle$,
which contributes to the flux of the turbulent
kinetic energy density.

In temperature-stratified turbulence, fluctuations of fluid temperature $\theta$
and velocity ${\bm u}$ are correlated due to a
non-zero turbulent heat flux, $\langle
\theta \, {\bm u} \rangle\not=\bm{0}$. Fluctuations of
temperature cause pressure fluctuations, which
result in fluctuations in the number density of
particles.
Increase of the pressure of the surrounding fluid
is accompanied by an accumulation of particles,
and the direction of the mean flux of particles
coincides with that of the turbulent heat flux.
The mean flux of particles is directed toward the
minimum of the mean temperature, and the
particles tend to be accumulated in this
region.

To demonstrate that the directions of the mean
flux of particles and the turbulent heat flux
coincide, we assume that the mean temperature
$\overline{T}_2$ at point $2$ is larger than the mean
temperature $\overline{T}_1$ at point $1$.
We consider two small control volumes {``a''} and
{``b''} located between these two points.
Let the direction of the local turbulent velocity in
volume {``a''} at some instant be
the same as the direction of the turbulent heat
flux $\langle \theta \, {\bm u} \rangle$ (i.e., along the
$x$--axis toward point $1$) and let the local
turbulent velocity in volume
{``b''}, at the same instant, be directed
opposite to the turbulent heat flux (i.e., toward
point $2$).

In temperature stratified turbulence
with a non-zero turbulent heat flux $\langle \theta \, {\bm u} \rangle$,
fluctuations of pressure $p$ and velocity ${\bm u}$
are correlated, and regions with a higher level
of pressure fluctuations have higher temperature
and velocity fluctuations.
Fluctuations of temperature $\theta$ and pressure $P$ in
volume {``a''} are positive because $\theta \, u_{x} > 0$, and negative in volume
{``b''}.
Fluctuations of particle number density $n'$ are
positive in volume {``a''} (because
particles are locally accumulated in the vicinity
of the maximum of pressure fluctuations, $d n' /
d t \propto - \overline{n}  \, (\tau_{\rm p} /\overline{\rho} )
\,\bec\nabla^2 P > 0$), and they are negative in
volume {``b''} (because there is an
outflow of particles from regions with low
pressure fluctuations). The flux of particles $n' \, {\bm V}$ is positive in volume {``a''}
(i.e., it is directed toward point $1$), and it
is also positive in volume {``b''}
(because both fluctuations of velocity and number
density of particles are negative in
volume {``b''}),
where ${\bm V}$ is the particle velocity
along the $x$-axis.
Therefore the mean flux of particles
$\langle n' \, {\bm V} \rangle$ is directed, as is the turbulent heat
flux $\langle \theta \, {\bm u}\rangle$, toward point~1.
This causes the formation of large-scale
inhomogeneous structures in the spatial
distribution of inertial particles in the
vicinity of the mean temperature minimum.

The increase of the rotation rate increases anisotropy of turbulence,
and fast rotation results in the effective pumping velocity is
in the plane perpendicular to the rotation axis.

\subsection{The effective pumping velocity ${\bm V}^{(4)}$}

Now we determine the effective pumping velocity  ${\bm V}^{(4)} = - (\tau_{\rm p} / \overline{\rho}) \, \langle \tau \,{\bm u} \,  (\hat{\bm \Omega} \cdot {\bm \nabla})^2 P \rangle$, that is given by Eq.~(\ref{CLG15}) in Appendix~\ref{appendix-A}.
For slow rotation, $\Omega_\ast^2 \ll 1$, the effective pumping velocity ${\bm V}^{(4)}$ is given by
\begin{eqnarray}
&& {\bm V}^{(4)}  = - {\mu_0 \, D_T \over 3} \, \ln{\rm Re} \, {\bm \nabla}  \ln \left(\overline{T} \, \overline{P}^{\, 1/\gamma-1}\right)  ,
\label{CLG16}
\end{eqnarray}
and for fast rotation, $1 \ll \Omega_\ast^2 \ll {\rm Re}$, the effective pumping velocity ${\bm V}^{(4)}$ is given by
\begin{eqnarray}
&& {\bm V}^{(4)}  = - {\mu_0 \, D_T \over 3 \varepsilon_u} \, \biggl[\ln\biggl({{\rm Re} \over \Omega_\ast^{2}} \biggr)
+ {2 \over 3} \biggr] \, {\bm \nabla}   \ln \left(\overline{T} \, \overline{P}^{\, 1/\gamma-1}\right) .
\label{CLG16}
\end{eqnarray}
Since for fast rotation $\Omega_\ast \gg 1$, the degree of anisotropy $\varepsilon_u$ of
turbulent velocity field is large $(\varepsilon_u \gg 1)$, the effective pumping velocity ${\bm V}^{(4)}$
vanishes.

\subsection{The effective pumping velocity ${\bm V}^{(5)}$}

Now we determine the effective pumping velocity ${\bm V}^{(5)} = - \langle \tau \,{\bm u} \,  (\hat{\bm \Omega} \cdot {\rm rot} \,  {\bm u}) \rangle$, that is given by Eq.~(\ref{CLG9}) in Appendix~\ref{appendix-A}.
For slow rotation, $\Omega_\ast^2 \ll 1$, the effective pumping velocity ${\bm V}^{(5)}$ is given by
\begin{eqnarray}
&& {\bm V}^{(5)}  = - {1 \over 4} \, \biggl[\hat {\bm \Omega} {\bm \times} {\bm \lambda}_\perp
+ {3 \over 2}\, \hat {\bm \Omega} {\bm \times} {\bm \nabla}_\perp -{4 \over 9}\, \hat {\bm \Omega} \, \Big(\hat {\bm \Omega} \cdot {\bm \lambda}\Big)\biggr] D_T ,
\nonumber\\
\label{CLG10}
\end{eqnarray}
and for fast rotation, $\Omega_\ast^2 \gg 1$, the effective pumping velocity ${\bm V}^{(5)}$ is given by
\begin{eqnarray}
&& {\bm V}^{(5)}  = D_T \left[\Omega_\ast {\bm \lambda}_\perp - {3 \over 2} \, \hat {\bm \Omega} {\bm \times} {\bm \lambda}_\perp \right].
\label{CLG11}
\end{eqnarray}

The mechanism for the appearance of an additional mean turbulent flux
of particles for fast rotation in the plane perpendicular to the rotation axis is as follows.
There are particle containing eddies with the vorticity $ {\bm \nabla} {\bm \times} {\bm u}$
which is parallel (the left-handed eddies) and antiparallel (the right-handed eddies)
to the global rotation $ {\bm \Omega}$.
The fluid density stratification in the plane perpendicular to the rotation axis
brakes a symmetry between number of the left-handed and right-handed eddies.
This results in appearance of a non-zero effective pumping velocity
${\bm V}^{(5)} = - \langle \tau \,{\bm u} \,  (\hat{\bm \Omega} \cdot {\rm rot} \,  {\bm u}) \rangle$
of particles in the direction of the density stratification.

\begin{figure}
\centering
\includegraphics[width=8.0cm]{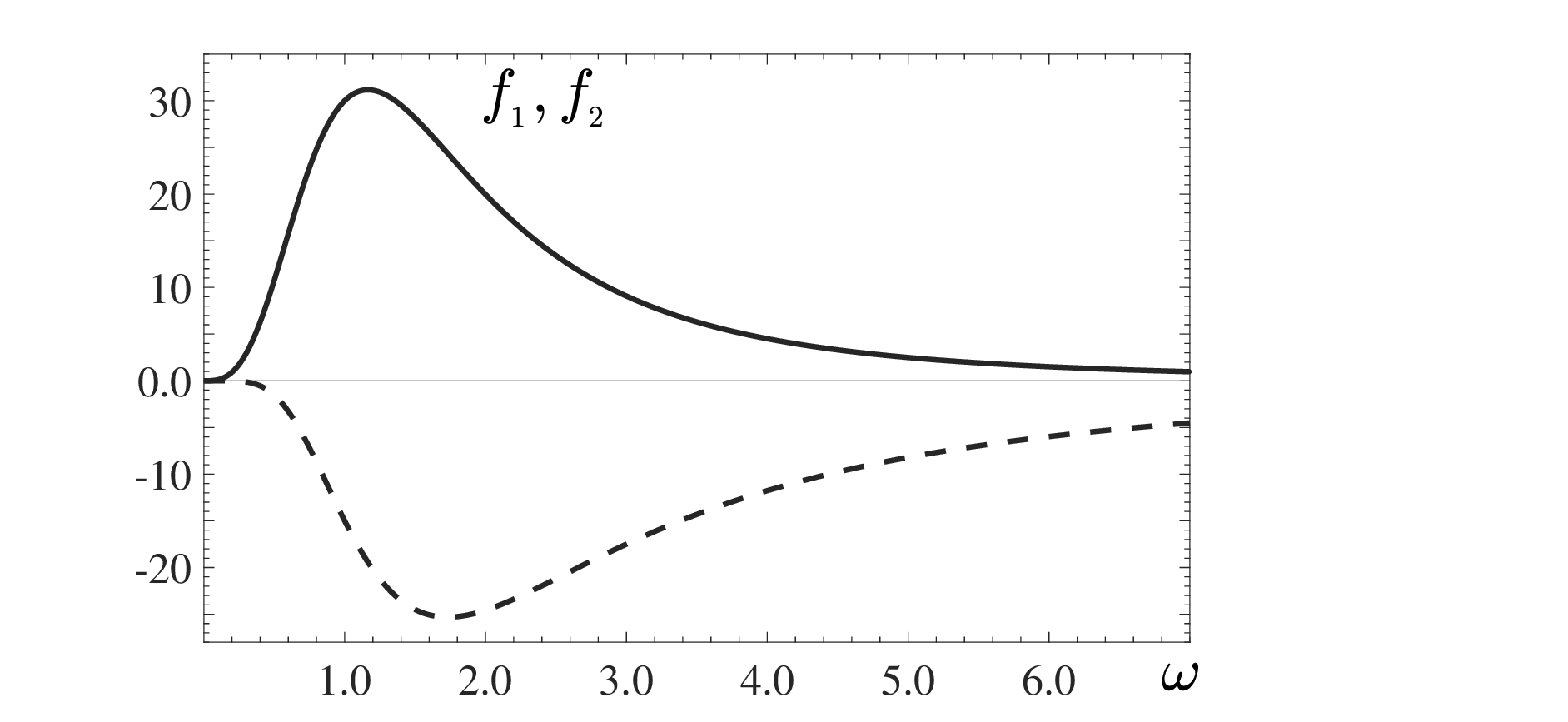}
\caption{\label{Fig1} The functions $f_1(\omega)$ (solid) and $f_2(\omega)$ (dashed)
for $\Omega_\ast =60$.
}
\end{figure}

For fast rotation $(\Omega_\ast^2 \gg 1)$,  the total effective pumping velocity
is given by
\begin{eqnarray}
&& {\bm V}^{(\rm eff)} =  f_1(\omega) \, D_T \, {\bm \lambda}_\perp
+ f_2(\omega) \, D_T \, \hat {\bm \Omega} {\bm \times} {\bm \lambda}_\perp ,
\label{CLS11}
\end{eqnarray}
where
\begin{eqnarray}
f_1(\omega) = 2 \Omega_\ast \, \omega^3 \, (1 +3 \omega^2) \, (1 + \omega^2)^{-4} ,
\label{CF1}
\end{eqnarray}

\begin{eqnarray}
f_2(\omega) = - 4 \Omega_\ast \, \omega^6  \, (1 + \omega^2)^{-4} ,
\label{CF2}
\end{eqnarray}
In Fig.~\ref{Fig1} we show the functions $f_1(\omega)$ and $f_2(\omega)$ entering to the total effective pumping velocity~(\ref{CLS11})
for fast rotating turbulence ($\Omega_\ast =60$).
In the next section, we analyse the behaviour of the total effective pumping velocity~(\ref{CLS11})
for a fast rotating turbulence.

\section{Applications to formation of planetesimals in accretion protoplanetary
discs}

The analyzed effects are important in astrophysical turbulence,
e.g., formation of planetesimals (progenitors of planets) in accretion protoplanetary
discs \cite{BR98,EKR98,PPK11,HUB16,HOP16,HOP16b}.
Small--scale planetesimals are formed from grains and
dust in the gaseous protostellar discs or the solar nebula due to
coagulation. Inertia of particles advected by turbulent rotating
fluid flow causes formation of large-scale
inhomogeneities of particles distribution.

The typical parameters of the protosolar nebula are \cite{BR98,PPK11,HOP16,HOP16b}:
the angular velocity $\Omega \sim 2 \times 10^{-7} \, r_{_{\rm AU}}^{-3/2} \, {\rm s}^{-1}$
(where $r_{_{\rm AU}}$ is the radial coordinate
measured in the astronomical units $L_{_{\rm AU}}=1.5 \times 10^{13}$ cm);
the sound speed $c_{\rm s} = 6.4 \times 10^{4} \, r_{_{\rm AU}}^{-3/14}$ cm/s;
the integral scale of turbulence $\ell_0 = \sqrt{\alpha} \, c_{\rm s}/\Omega $;
the turbulent velocity $u_0 = \alpha \, c_{\rm s}$;
the turbulent time $\tau_0=\ell_0/u_0 = (\sqrt{\alpha} \Omega)^{-1}$
and the turbulent diffusion coefficient $D_T= \ell_0 \, u_0 / 3$ varies from $2 \times 10^{11} \, r_{_{\rm AU}}^{15/14}$ cm$^2$/s to $10^{13} \, r_{_{\rm AU}}^{15/14}$ cm$^2$/s.
The kinematic viscosity $\nu=c_{\rm s} \, \lambda_{\rm mfp} /2 = 1.6 \times 10^{5} \, r_{_{\rm AU}}^{18/7}$ cm$^2$/s,
so the Reynolds number ${\rm Re} = \ell_0 \, u_0 / \nu$ varies from
$3 \times 10^{6} \, r_{_{\rm AU}}^{-3/2}$ to $10^{8} \, r_{_{\rm AU}}^{-3/2}$.
Here $\lambda_{\rm mfp} = 5 \, r_{_{\rm AU}}^{39/14}$ cm is the mean-free path of the gas molecules,
$\alpha$ varies from $10^{-3}$ to $10^{-2}$.
Therefore, the parameter $\Omega_\ast = 4 \Omega \tau_0 = 4/\sqrt{\alpha}$ varies from
$40$ to $120$. This implies that turbulence in the protosolar nebula is a fast rotating turbulence.

The stoping time $\tau_{\rm p}$ describing particle and fluid interaction
depends on the ratio of the mean-free path $\lambda_{\rm mfp}$ of the gas molecules to the particle radius $a_{\rm p}$.
In the Epstein regime ($\lambda_{\rm mfp} > a_{\rm p}$), the stoping time is
$\tau_{\rm p} = \rho_{\rm p} \, a_{\rm p}/ (\overline{\rho} \, c_{\rm s}) = 2 \times 10^{4} \, r_{_{\rm AU}}^{18/7} a_{\rm p}$,
where the particle radius $a_{\rm p}$ is measured in cm and $\tau_{\rm p}$ is measured in s.
Therefore, the parameter $\omega = 2 \Omega \tau_{\rm p}$ is $\omega = 10^{-2} r_{_{\rm AU}}^{15/14} a_{\rm p}$.
In the Stokes regime ($\lambda_{\rm mfp} \ll a_{\rm p}$), the stoping time is
$\tau_{\rm p} = 2 \rho_{\rm p} \, a_{\rm p}^2/(9 \overline{\rho}\, \nu) = 2 \times 10^{3} \, r_{_{\rm AU}}^{3/14} a_{\rm p}^2$.
Therefore, the parameter $\omega = 10^{-3} r_{_{\rm AU}}^{-9/7} a_{\rm p}^2$.

\begin{figure}
\centering
\includegraphics[width=8.0cm]{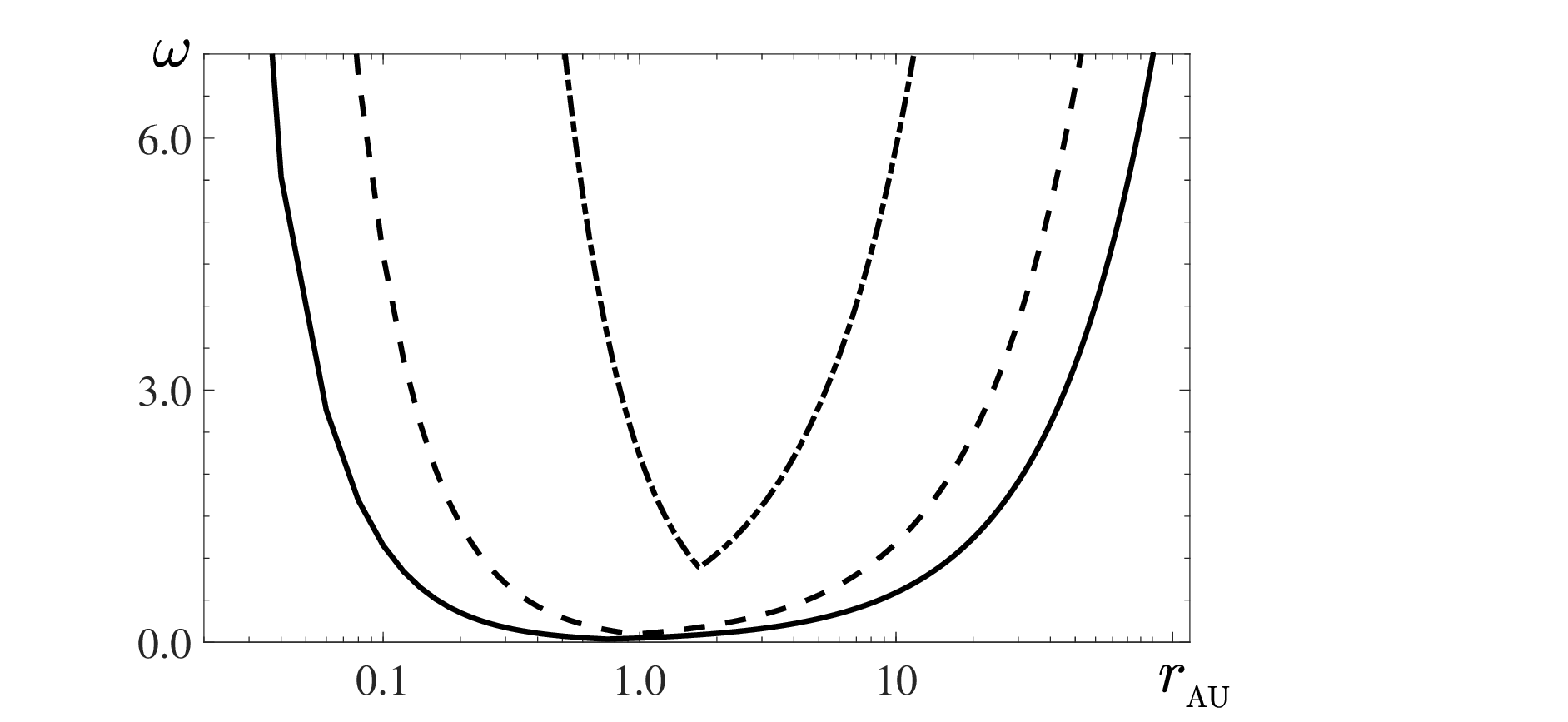}
\caption{\label{Fig2}
The radial profiles of the parameter $\omega(r_{_{\rm AU}})$ for different particle radius: $a_{\rm p} = $ 5 cm (solid), 10 cm (dashed); 50 cm (dashed-dotted). Here $r_{_{\rm AU}}$ is the radial coordinate
measured in the astronomical units $L_{_{\rm AU}}=1.5 \times 10^{13}$ cm.
}
\end{figure}

In Fig.~\ref{Fig2} we show the radial profiles of the parameter $\omega(r_{_{\rm AU}})$ for different particle radius: $a_{\rm p} = $ 5 cm (solid), 10 cm (dashed); 50 cm (dashed-dotted). The decreased function $\omega(r_{_{\rm AU}})$ corresponds to the Stokes regime, while the increased function $\omega(r_{_{\rm AU}})$ corresponds to the Epstein  regime.
To plot Fig.~\ref{Fig2}, we use  Eq. (1) from \cite{HOP16b} for $\tau_{\rm p}$ with the smooth transition
between the Stokes and the Epstein regimes.
In particular, the case $a_{\rm p} \leq 9 \lambda_{\rm mfp}/4$ corresponds to the Epstein regime, while
for  $a_{\rm p} > 9 \lambda_{\rm mfp}/4$ the Stokes regime
describes $\tau_{\rm p}$, where $\lambda_{\rm mfp} = 5 \, r_{_{\rm AU}}^{39/14}$ cm.

\begin{figure}
\centering
\includegraphics[width=8.0cm]{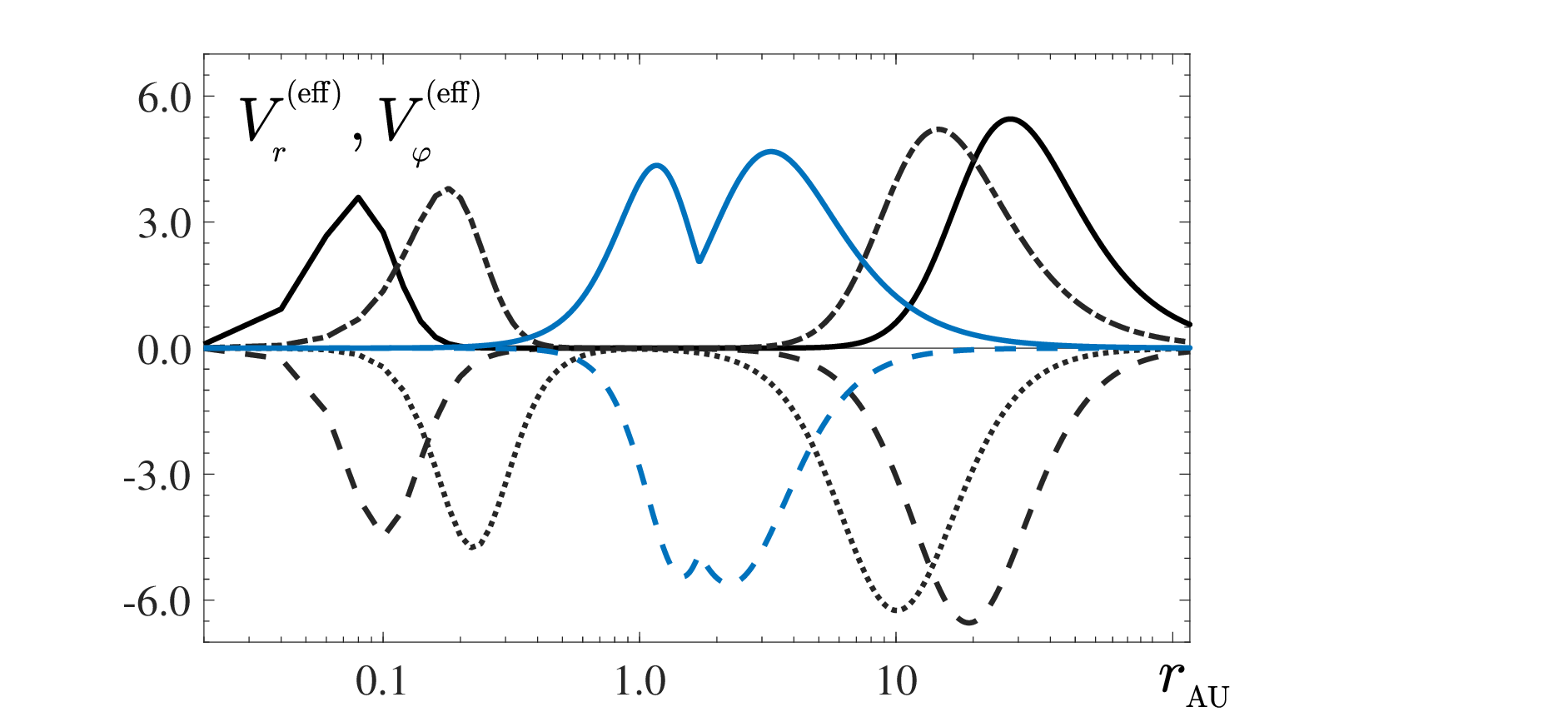}
\caption{\label{Fig3}
The radial profiles of the azimuthal $V_\varphi^{(\rm eff)}(r_{_{\rm AU}})$ component of the effective velocity for different particle radius: $a_{\rm p} = $ 5 cm (solid), 10 cm (dashed-dotted); 50 cm (solid blue);
and of the radial component $V_r^{(\rm eff)}(r_{_{\rm AU}})$ of the effective velocity for different particle size $a_{\rm p} = $ 5 cm (dashed), 10 cm (dotted); 50 cm (dashed blue) for $\Omega_\ast =60$ (that corresponds to $\alpha=1/225$).
The velocity is measured in cm/s.
}
\end{figure}

The mean fluid density is $\meanrho = 2 \times 10^{-9} \, r_{_{\rm AU}}^{-11/4}$ g/cm$^3$,
the mean fluid temperature is $\meanT = 280 \, r_{_{\rm AU}}^{-1/2}$ K,
the parameter $\mu_0 \, D_T = (2/3 \gamma) \, \tau_{\rm p} \, c_{\rm s}^2$,
the scale $H_{\rm g}=c_{\rm s}/\Omega=3 \times 10^{11} \, r_{_{\rm AU}}^{9/7}$ cm.
Note that in the radial direction the gravity force acting to the particles is compensated by the centrifugal force.
This is the reason why these forces are not included in Eqs.~(\ref{CLD1})--(\ref{CLD2}).
In Fig.~\ref{Fig3} we plot the radial profiles of the azimuthal $V_\varphi^{(\rm eff)}(r_{_{\rm AU}})$ and radial $V_r^{(\rm eff)}(r_{_{\rm AU}})$ components of the total effective velocity  for different particle radius $a_{\rm p} = $ from 5 cm to 50 cm
for fast rotating turbulence ($\Omega_\ast =60$).
The effective pumping velocity components for fast rotating turbulence are determined by Eq.~(\ref{CLS11}),
where ${\bm \lambda}_\perp$ is directed in the radial direction.
These radial profiles have two maxima corresponding to different regimes of the particle--fluid interactions: at the small radius it is the Stokes regime, while at the larger radius it is the Epstein regime. With the decrease the particle radius, the distance between the maxima increases.
This implies that smaller-size particles are concentrated nearby the central body of the accretion disk, while larger-size particles are accumulated far from the central body.

\begin{figure}
\centering
\includegraphics[width=8.0cm]{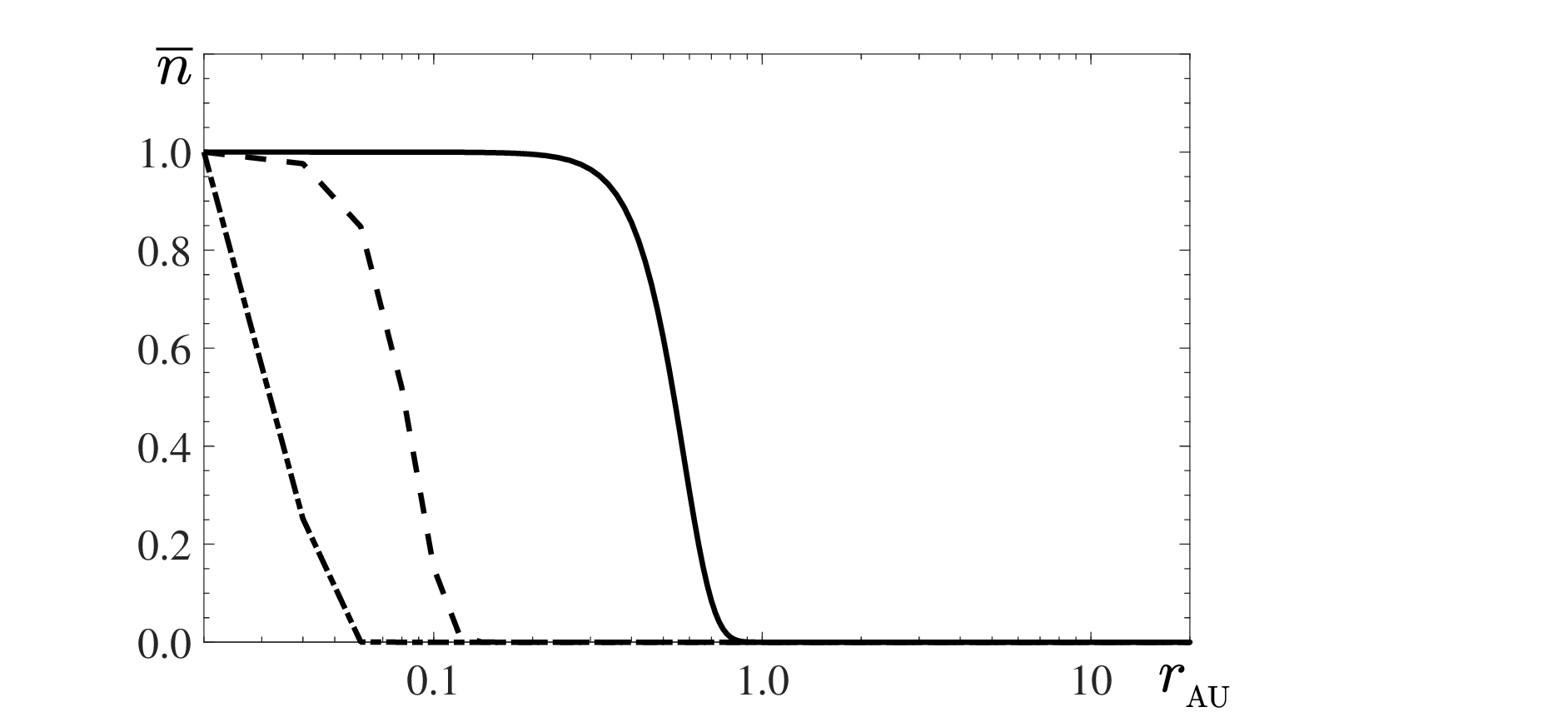}
\caption{\label{Fig4} The equilibrium radial profile of the normalised mean particle number density $\meanN(r_{_{\rm AU}})$ for different particle radius: $a_{\rm p} = $ 5 cm (dashed-dotted), 10 cm (dashed); 50 cm (solid) and for $\Omega_\ast =60$.
}
\end{figure}

In Fig.~\ref{Fig4} we plot the equilibrium radial profile of the mean particle number density
$\meanN(r_{_{\rm AU}})$ for different particle radius $a_{\rm p}$ from 5 cm to 50 cm
for fast rotating turbulence ($\Omega_\ast =60$).
The equilibrium radial profile of the mean particle number density
$\meanN(r_{_{\rm AU}})$ is determined the steady-state solution of Eq.~(\ref{CLA2}):
\begin{eqnarray}
\meanN(r_{_{\rm AU}}) =\meanN_\ast \, \exp \left( - \int_{r_{\rm min}}^{r_{_{\rm AU}}} {V_r^{(\rm eff)} (r') \over D_T(r') }
\, dr' \right) .
\label{PR10}
\end{eqnarray}
It follows from Fig.~\ref{Fig4} that the size of particle cluster increases with the particle radius.
Note that the material density of particles $\rho_{\rm p}$ as well as their radius $a_{\rm p}$ affect the particle stoping time.
With increase of  $\rho_{\rm p}$ and $a_{\rm p}$ the particle stoping time increases.
The characteristic formation time of large-scale particle clusters is about $\tau_{\rm dyn} = r / V_r^{(\rm eff)} \sim 10^5$--$10^6$ years,
where $r_{_{\rm AU}} = 1$--$10$ (measured in the astronomical units $L_{_{\rm AU}}$) and $V_r^{(\rm eff)} \sim 5$ cm/s.
Note that the turbulent diffusion time is about $\tau_{\rm diff} = r^2/D_T \sim (0.4$--$3) \times 10^7$ years.
The turbulent diffusion affects  the life-time of the large-scale clusters, and $\tau_{\rm diff}$ is much larger than the characteristic formation time of large-scale particle cluster ($\sim \tau_{\rm dyn}$).

\section{Discussion and conclusions}

A theory of  large-scale clustering of inertial particles
(in scales which are much larger than the turbulent integral scale)
in a rotating density stratified or inhomogeneous turbulence is developed.
The large-scale particle clustering
is characterised by the effective pumping velocity
in a turbulent flux of particles, which
for a fast rotation is localized in the plane perpendicular to rotation axis
along the fluid density stratification.
This causes formation of large-scale
inhomogeneities in particle spatial distribution.
The developed theory of the large-scale particle clustering has been applied
for explanation of the formation of planetesimals (progenitors of planets)
in accretion protoplanetary discs.

We apply mean-field theory to describe large-scale clustering phenomena
in rotating and stratified turbulence.
The mean-field theory is valid when the characteristic spatial (and/or  temporal)
scales of the mean-field variations are much larger than the integral turbulence scale (and/or  turbulence correlation time).
The applicability of a mean-field theory in fast rotating and stratified turbulence based on the $\tau$ approach has been discussed
in Ref.~\cite{RK19}, where comparison with results of direct numerical simulations (see Refs.~\cite{PMH11,MPH11})
has been performed.
In these papers a generation of a large-scale vorticity in a
fast rotating density stratified turbulent convection has been studied.

The theory \cite{RK19} predicts the threshold in the Coriolis number requires for the
generation of the large-scale vorticity observed in DNS \cite{PMH11,MPH11}. The critical Coriolis
number should be much larger than 1. The derived mean-field
equations describe formations of both, cyclonic and anticyclonic
large-scale vortices in the kinematic (linear) stage of
the instability. As in the DNS, the theory \cite{RK19} predicts the similar behavior
of the mean entropy or temperature inside cyclonic and
anticyclonic vortices: for the mode with the dominant vertical mean vorticity, the mean
entropy is decreased inside the cyclonic vortices and increased
inside the anticyclonic vortices in agreement with DNS.
Detailed validation of the developed mean-field theory of
large-scale clustering in rotating density-stratified turbulence
in numerical simulations is a subject of a future separate studies.

To derive equations for the rotational contribution to the Reynolds stress
and the turbulent heat flux in a rotating density stratified turbulence,
we apply the spectral $\tau$ approach (see Section~II and Appendix B).
The $\tau$ approach is valid for large fluid and magnetic Reynolds numbers and for large
P\'{e}clet numbers.
In this case turbulence is strong, and the relaxation time $\tau_r$ in Eq.~(\ref{CCB6}) can be identified
with the scale-dependent turbulent time $\tau(k)$ (see Appendix~B).
However, the $\tau$ approach does not work to study intermittency of scalar and magnetic fluctuations.

The $\tau$ approach reproduces many well-known phenomena found
by other methods in turbulent transport of particles, temperature and
magnetic fields, in turbulent convection and stably stratified
turbulent flows (see detailed discussion in Ref.~\cite{RI21}, and references therein).
In particular, in turbulent transport of particles, the $\tau$ approximation yields correct
formulae for turbulent diffusion, turbulent thermal diffusion
and turbulent barodiffusion obtained by other methods.
The phenomenon of turbulent thermal diffusion
(a nondiffusive streaming of particles in the direction of
the turbulent heat flux), has been studied using the stochastic calculus (the path
integral approach), the quasi-linear approach, the direct interaction approximation (DIA)
and the $\tau$ approach (see Ref.~\cite{RI21}, and references therein).
The phenomenon of turbulent thermal diffusion and large-scale particle clustering has been already detected in laboratory experiments
in  non-rotating stably and unstably temperature
stratified turbulence \cite{BEE04,EEKR06,EKRL22,SKRL22,EKRL23},
and in direct numerical simulations (DNS) in non-rotating density-stratified turbulence
\cite{HKRB12,BRK12,RKB18}.
The numerical and experimental results are
in a good agreement with the theoretical studies performed by means
of different approaches, including the $\tau$ approach.

The detailed verification of the $\tau$ approach in DNS of turbulent transport of passive scalar has been
performed in \cite{BKM04}. In particular, the results
on turbulent transport of passive scalar obtained using DNS have been
compared with that obtained using a closure model based on the
$\tau$ approach. The numerical and analytical results are in a
good agreement.

The $\tau$ approach reproduces the
well-known $k^{-7/3}$-spectrum of anisotropic velocity fluctuations
in a sheared turbulence \cite{EKRZ02,EKRZ06}.
This spectrum was previously found in analytical \cite{L67}, numerical \cite{IYK02},
laboratory studies \cite{SV94} and was observed in the atmospheric
turbulence \cite{WC72}.
In the turbulent boundary layer problems, the
$\tau$ approach yields correct expressions for turbulent viscosity,
turbulent thermal conductivity and the classical turbulent heat flux.
This approach also describes the counter-wind heat flux
and the Deardorff's heat flux in convective turbulence \cite{EKRZ02,EKRZ06}.
These phenomena have been previously studied using different analytical approaches
(see Ref.~\cite{RI21}, and references therein).

In magnetohydrodynamics, the $\tau$ approach reproduces many
well-known phenomena found by different methods, e.g., the
$\tau$ approach yields correct formulae
for the $\alpha$-effect, the effective pumping velocities,
and the turbulent magnetic diffusion
in rotating and density stratified turbulence
(see Ref.~\cite{RI21}, and references therein).
These results also have been confirmed using DNS
in forced turbulence as well as in rotating density-stratified turbulence
\cite{BRK12,BKR13}.
In high-energy physics, e.g., in chiral magnetohydrodynamics, DNS reproduce
many results predicted using the $\tau$ approach \cite{RKB17,BRK17,SRB18,SRB22,SRB24}.

Comparisons of theoretical predictions and DNS performed
for various turbulence problems (where large-scale effects are caused by turbulence) show that often
the separation of scales characterised by the ratio of typical scale of mean-field variations
to the integral scale of turbulence should be in DNS in the interval from 3 to 5.
This allows to observe in DNS the large-scale effects
caused by turbulence.
However, there is a case when the scale separation ratio should be much larger.
For instance, the negative effective magnetic pressure instability (NEMPI)  \cite{BKKR11,BRK16},
which results in the formation of the large-scale magnetic flux tubes,
requires the scale separation ratio in the interval from 15 to 30.
This allows to observe this instability in DNS \cite{BKKR11,BRK16}.

In the present study we consider large-scale clustering in a low-Mach-number turbulence.
These conditions are relevant for applications to the formation of planetesimals
in accretion protoplanetary discs.
The weak correlation between fluid density and velocity fluctuations at low-Mach-number turbulence
is discussed in Ref.~\cite{CH02}.
The effects of compressibility (not small Mach numbers) on the turbulent diffusivity and the effective pumping velocity
of particles for non-rotating density stratified or inhomogeneous turbulence are
discussed in Section 2.5 in \cite{RI21}, see references therein. The theoretical
predictions of the compressibility effects on turbulent transport of particles are
in agreement with DNS results for non-rotating density stratified
or inhomogeneous turbulence \cite{BRK12,RKB18}.
The effects of compressibility on the large-scale particle clustering in a rotating density stratified
and inhomogeneous turbulence is the subject of future separate studies.

In the developed theory of large-scale particle clustering,
the stopping time $\tau_{\rm p}$ which describes the fluid - particle interaction regimes,
inters only in the parameter $\omega=2 \Omega \tau_{\rm p}$,
and the theory is developed for arbitrary parameter $\omega$.
In the applications of the analyzed effects for explanation of formation of planetesimals in protoplanetary discs,
we consider two different fluid - particle interaction regimes.
In particular, the stoping time $\tau_{\rm p}$ depends on the ratio of the mean-free path $\lambda_{\rm mfp}$
of the gas molecules to the particle radius $a_{\rm p}$.
When $\lambda_{\rm mfp} / a_{\rm p} > 1$, the Epstein regime of the fluid - particle interaction
describes $\tau_{\rm p}$, while for $\lambda_{\rm mfp} / a_{\rm p} \ll 1$, the
Stokes regime describes $\tau_{\rm p}$. When the ratio $\lambda_{\rm mfp} / a_{\rm p} < 1$, and it is not small,
there is a smooth transition between these regimes.
For instance, to plot Fig.~\ref{Fig2} for the radial profiles of the parameter $\omega(r_{_{\rm AU}})$,
we use  Eq. (1) from \cite{HOP16b} for $\tau_{\rm p}$ with the smooth transition
between the Stokes and the Epstein regimes (see Section IV).
Another way is to use Eq. (10) from \cite{BD10} for $\tau_{\rm p}$ with the smooth transition between different regimes.
However, this only affects the parameter $\omega$.

In this paper we analyse the reasons for large-scale clustering
caused by the turbulent effects, which are described in terms of the effective pumping velocity
in the mean-field equation for the mean particle number density.
The derived pumping velocities are independent of boundary conditions.
These pumping velocities arise due to small-scale turbulence in the presence of rotation
and stratification.

To determine the characteristics of the large-scale clusters, like
dimensions (the characteristic length and width) and shape (aspect ratio) of the large-scale clusters,
and time evolution of clusters,
one needs to solve the derived nonlinear mean-field equations which include
the derived effective pumping velocities.
This can be done by the mean-field numerical simulations (MFS)
which is the subject of future separate comprehensive studies.
The subsequent particle distributions in the clusters obtained in MFS may depend
on boundary conditions.
Note also that a mean-field theory is able to take into account the back-reactions of particles on turbulence
which are important when the mass-loading parameter $m_{\rm p} \overline{n}/\overline{\rho}$ inside the cluster is of the order of
or larger than 1.
These nonlinear effects can be studied in MFS.

In the developed theory there are two key parameters: $\omega = 2 \tau_p \Omega$
and $\Omega_\ast = 4 \Omega \tau_0$.
The stopping time of particles $\tau_{\rm p} \ll \tau_0$ is the smallest time in the system,
and the theory is valid for large Reynolds numbers and for large
P\'{e}clet numbers.
The developed theory can be  also applied to large-scale clustering of dust in accretion disks in binary stellar system
and in young galactic rotating accretion disks.

\begin{acknowledgments}
I.R. would like to thank the Isaac Newton Institute for Mathematical Sciences,
Cambridge, for support and hospitality during the programme "Anti-diffusive dynamics:
from sub-cellular to astrophysical scales", where this work
was initiated.
I.R.  acknowledges the discussions with some participants of the Nordita Scientific Program
on ‘Stellar Convection: Modelling, Theory and Observations’, Stockholm (September 2024),
which is partly supported by NordForsk.
\end{acknowledgments}

\bigskip
\noindent
{\bf DATA AVAILABILITY}
\medskip

The data that support the findings of this study are available from the corresponding author
upon reasonable request.

\medskip
\noindent
{\bf AUTHOR DECLARATIONS}

\noindent
{\bf Conflict of Interest}

The authors have no conflicts to disclose.

\appendix

\section{Total effective pumping velocity of particles}
\label{appendix-A}

In this Appendix, we derive the total effective pumping velocity
${\bm V}^{(\rm eff)} = - \langle \tau \,{\bm V} \,  {\rm div} \,
{\bm V} \rangle$ for arbitrary values of parameter $\omega$.
To determine ${\rm div} \, {\bm V}$ [given by Eq.~(\ref{CLD17})],
we assume that the parameter $\omega$ and the Stokes time $\tau_{\rm p}$
are independent parameters.
Using Eq.~(\ref{CLD4}), we find $d{\bm V} / d t$
for small Stokes time applying a method of iterations. This yields
\begin{eqnarray}
&& - \tau_{\rm p} \, {d{\bm V} \over d t}= {\omega \over 1 + \omega^2} \,
\biggl[\hat{\bm \Omega} {\bm \times} {\bm u}+ \omega \, \Big({\bm u} - \hat{\bm \Omega} \, (\hat{\bm \Omega} \cdot {\bm u})\Big)
\nonumber\\
&& \quad + {\tau_{\rm p} \over \rho} \, \Big(\omega^{-1}\, {\bm \nabla} + \omega \,\hat{\bm \Omega} \, (\hat{\bm \Omega}  \cdot  {\bm \nabla}) - \hat{\bm \Omega} {\bm \times} {\bm \nabla} \Big) P \biggr] + {\rm O}\left(\tau_{\rm p}^2\right) ,
\nonumber\\
\label{CLD14}
\end{eqnarray}
and
\begin{eqnarray}
&& - \tau_{\rm p} \, B_{mi} \, {dV_i \over d t}= {\omega^2 \over 1 + \omega^2} \,
\biggl[{\bm u} - \omega\, \hat{\bm \Omega} {\bm \times} {\bm u} - \hat{\bm \Omega} \, (\hat{\bm \Omega} \cdot {\bm u})
\nonumber\\
&& \quad - {\tau_{\rm p} \over \rho} \, \Big({\bm \nabla} - (2 + \omega^2) \, \hat{\bm \Omega} \, (\hat{\bm \Omega}  \cdot  {\bm \nabla}) + \omega^{-1}\, \hat{\bm \Omega} {\bm \times} {\bm \nabla} \Big) P \biggr]_m
\nonumber\\
&& \quad+ {\rm O}\left(\tau_{\rm p}^2\right) ,
\label{CLD18}
\end{eqnarray}
where  $\hat{\bm \Omega} ={\bm \Omega} / \Omega$ is the unit vector
and we use Eq.~(\ref{CLD2}) for large Reynolds numbers (which allows us
to neglect small term proportional to the kinematic viscosity).
Equations~(\ref{CLD17}) and Eqs.~(\ref{CLD14})--(\ref{CLD18}) allow us to find the compressibility
of particle velocity field characterised by ${\rm div} \, {\bm V}$:
\begin{eqnarray}
&&{\rm div} \, {\bm V} = {\rm div} \, {\bm u} + {\tau_{\rm p} \over \rho} \, (\Delta + {\bm \lambda}  \cdot  {\bm \nabla}) P
+ {\omega^2 \over (1 + \omega^2)^2}
\nonumber\\
&& \times \biggl\{2\omega \,  \hat{\bm \Omega} \cdot  {\rm rot} \, {\bm u}+ (1- \omega^2) \, {\rm div} \, {\bm u}_\perp
- {\tau_{\rm p} \over \rho} \,\biggl[ (3 + \omega^2) \, (\Delta_\perp
\nonumber\\
&& + {\bm \lambda}_\perp  \cdot  {\bm \nabla}) + 2\omega^{-1}\,   (\hat{\bm \Omega} {\bm \times} {\bm \lambda}_\perp ) \cdot {\bm \nabla} \biggr] P \biggr\}
 + {\rm O}\left(\tau_{\rm p}^2\right) ,
\label{CLD6}
\end{eqnarray}
where ${\bm \lambda} = - \meanrho^{\, -1} \, {\bm \nabla} \meanrho$.
For the derivation of Eqs.~(\ref{CLD14})--(\ref{CLD6}), we used the following identities:
$B_{mi} \, \nabla_i =\omega^2\, [\hat {\bm \Omega} (\hat {\bm \Omega} \cdot {\bm \nabla}) - \omega^{-1} \, \hat {\bm \Omega} {\bm \times} {\bm \nabla}]_m$,
\begin{eqnarray}
&& - \tau_{\rm p} \, {\rm div} \,  \left({d{\bm V} \over d t} \right) = {1 \over 1 + \omega^2} \,  \left(\delta_{mi} + B_{mi}\right)  \,
\biggl[{\tau_{\rm p} \over \rho} (\nabla_m
\nonumber\\
&& \quad + \lambda_m) \nabla_i  P - \omega \, \varepsilon_{ipq} \hat \Omega_q \, \nabla_m  u_p
\biggr] + {\rm O}\left(\tau_{\rm p}^2\right) ,
\label{CLD5}
\end{eqnarray}

\begin{eqnarray}
B_{mi} \, \nabla_i  \nabla_m P =  \omega^2 \, \left(\hat{\bm \Omega} \cdot {\bm \nabla}\right)^2 P ,
\label{CLD7}
\end{eqnarray}

\begin{eqnarray}
&& B_{mi} \, (\nabla_i  P) \, \lambda_m = \omega \, \biggl[\hat{\bm \Omega} \cdot ({\bm \lambda}_\perp {\bm \times}  {\bm \nabla})
+ \omega \,  (\hat{\bm \Omega} \cdot {\bm \lambda}) \, (\hat{\bm \Omega} \cdot {\bm \nabla})\biggr] P
\nonumber\\
\label{CLD8}
\end{eqnarray}

\begin{eqnarray}
B_{mi} \, \varepsilon_{ipq} \hat{\Omega}_q \, u_p = \omega \, \left[\hat{\Omega}_m \, \left(\hat{\bm \Omega} \cdot {\bm u}\right) - u_m\right] ,
\label{CLD12}
\end{eqnarray}

\begin{eqnarray}
B_{mi} \, \varepsilon_{ipq} \hat{\Omega}_q \, \nabla_m  u_p = - \omega  \, {\rm div} \, {\bm u}_\perp .
\label{CLD9}
\end{eqnarray}
This yields also
\begin{eqnarray}
&& - \tau_{\rm p} \, {\rm div} \,  \left({d{\bm V} \over d t} \right) = {\tau_{\rm p} \over \rho} \, \biggl[\Delta + {\bm \lambda}  \cdot  {\bm \nabla} - {\omega^2 \over 1 + \omega^2} \biggl(\Delta_\perp
\nonumber\\
&& \quad  + {\bm \lambda}_\perp  \cdot  {\bm \nabla} - \omega^{-1} \, \hat{\bm \Omega} \cdot ({\bm \lambda}_\perp {\bm \times} {\bm \nabla})
  \biggr) P  - {\rm div} \, {\bm u}_\perp
\nonumber\\
&& \quad + \omega^{-1} \, \hat{\bm \Omega} \cdot  {\rm rot} \, {\bm u}\biggr]+ {\rm O}\left(\tau_{\rm p}^2\right) ,
\label{CLD10}
\end{eqnarray}
and
\begin{eqnarray}
&& - \tau_{\rm p} \, \nabla_m \,  \left(B_{mi} \, {dV_i \over d t} \right) = {\omega^2 \over 1 + \omega^2} \, \biggl\{
{\tau_{\rm p} \over \rho} \, \biggl[(1+\omega^2) \,  (\Delta
\nonumber\\
&& + {\bm \lambda}  \cdot  {\bm \nabla})  - (2 + \omega^2) \,  (\Delta_\perp
+ {\bm \lambda}_\perp  \cdot  {\bm \nabla}) + \omega^{-1} (\hat{\bm \Omega} {\bm \times} {\bm \lambda}) \cdot {\bm \nabla} \biggr] P
\nonumber\\
&& + {\rm div} \, {\bm u}_\perp + \omega \, \hat{\bm \Omega} \cdot  {\rm rot} \, {\bm u}\biggr\} + {\rm O}\left(\tau_{\rm p}^2\right) .
\label{CLD11}
\end{eqnarray}
Equations~(\ref{CLD4}) and Eqs.~(\ref{CLD14})--(\ref{CLD18}) yield the particle velocity field
\begin{eqnarray}
&&{\bm V}  = {1 \over (1 + \omega^2)^2} \, \biggl\{(1 +3 \omega^2) {\bm u}  - 2 \omega^3\, (\hat{\bm \Omega} {\bm \times} {\bm u})
\nonumber\\
&& \quad -(1-\omega^2) \, \omega^2 (\hat{\bm \Omega} \cdot {\bm u}) \, \hat{\bm \Omega} + {\tau_{\rm p} \over \rho} \, \Big[(1-\omega^2) {\bm \nabla}
\nonumber\\
&& \quad+ (3 + \omega^2)  \omega^2\, \hat{\bm \Omega} \,  (\hat{\bm \Omega} \cdot {\bm \nabla})  - 2 \omega(\hat{\bm \Omega} {\bm \times} {\bm \nabla})\Big] P\biggr\} .
\label{CLD20}
\end{eqnarray}

Equations~(\ref{CLD6}) and~(\ref{CLD20}) for $\omega^2 \ll 1$ read:
\begin{eqnarray}
&&{\rm div} \, {\bm V} = (1 + \omega^2) \, {\rm div} \, {\bm u} - \omega^2 \, \left(\hat{\bm \Omega} \cdot {\bm \nabla}\right)
 \left(\hat{\bm \Omega} \cdot {\bm u}\right)
 \nonumber\\
&& \; + {\tau_{\rm p} \over \rho} \, \biggl[(1 - 3 \omega^2) \, \Big(\Delta + {\bm \lambda} \cdot  {\bm \nabla} \Big)
 + 3 \omega^2 \, \Big[ \left(\hat{\bm \Omega} \cdot {\bm \nabla}\right)^2
 \nonumber\\
&& \; + \left(\hat{\bm \Omega} {\bm \times} {\bm \lambda}\right)
 \left(\hat{\bm \Omega} \cdot {\bm \nabla}\right)  \Big]
+ 2\omega \,   \left(\hat{\bm \Omega} {\bm \times} {\bm \lambda}_\perp\right) \cdot {\bm \nabla}\biggr] P
\nonumber\\
&& \; + {\rm O}\left(\tau_{\rm p}^2\right) ,
\label{CLD15}
\end{eqnarray}
and
\begin{eqnarray}
&& {\bm V} = (1 + \omega^2) \, {\bm u} - \omega^2 \hat{\bm \Omega}\, (\hat{\bm \Omega} \cdot {\bm u}) + {\tau_{\rm p} \over \rho} \, \Big[(1 - 3 \omega^2) \, {\bm \nabla}
\nonumber\\
&& \; +3 \omega^2 \, \hat{\bm \Omega}\, (\hat{\bm \Omega} \cdot {\bm \nabla})  - 2 \omega\,(\hat{\bm \Omega} {\bm \times} {\bm \nabla}) \Big] P + {\rm O}\left(\tau_{\rm p}^2\right) .
\label{CLD21}
\end{eqnarray}
By means of Eqs.~(\ref{CLD6}) and~(\ref{CLD20}), we determine the total effective pumping velocity:
${\bm V}^{(\rm eff)} = - \left\langle \tau \,{\bm V} \,  {\rm div} \,  {\bm V} \right\rangle$,
which is given by  Eqs.~(\ref{CLS0}), where
the effective pumping velocity
${\bm V}^{(1)} = - \left\langle \tau \,{\bm u} \,  {\rm div} \,  {\bm u} \right\rangle$ is
\begin{eqnarray}
&& {\bm V}^{(1)} = - {1 \over 1 +\varepsilon_u} \, \biggl\{2 A_{1}^{(1)}(\Omega_\ast) \, {\bm \lambda}
+ \left[A_{2}^{(1)}(\Omega_\ast) + {3 \over 4} \,  \varepsilon_u \right] \, {\bm \lambda}_\perp
\nonumber\\
&& \quad
- \Omega_\ast \, A_{3}^{(2)}(\Omega_\ast) \biggl[ \hat{\bm \Omega} {\bm \times} \biggl({\bm \lambda}_\perp - {1 \over 2} {\bm \nabla}_\perp\biggr) \biggr] \biggr\} D_{T} .
\label{CLB10}
\end{eqnarray}
The functions $A_{n}^{(m)}(\Omega_\ast)$ are given in Appendix~\ref{appendix-C}.
The effective pumping velocity ${\bm V}^{(2)} = - \langle \tau \,{\bm u}  (\hat{\bm \Omega} \cdot {\bm \nabla})
(\hat{\bm \Omega} \cdot {\bm u}) \rangle$ is given by:
\begin{eqnarray}
&& {\bm V}^{(2)} = {1 \over 2(1 +\varepsilon_u)} \, \biggl[\Big(A_{1}^{(1)}(\Omega_\ast) - A_{2}^{(1)}(\Omega_\ast)  \Big)  \hat {\bm \Omega} \Big(\hat {\bm \Omega} \cdot{\bm \nabla} \Big)
\nonumber\\
&&  \quad + 2A_{3}^{(1)}(\Omega_\ast) \, {\bm \lambda}_\perp
+ 2 \, \Omega_\ast \,  \biggl(A_{2}^{(2)}(\Omega_\ast) + {4 \over 5} C_{1}^{(2)}(\Omega_\ast)  \biggr)
\nonumber\\
&& \quad \times \Big[\hat{\bm \Omega} {\bm \times}  \Big({\bm \lambda}_\perp - {1 \over 2} {\bm \nabla}_\perp\Big)\Big] \biggr] D_{T} .
\label{CLL1}
\end{eqnarray}
The functions $A_{n}^{(m)}(\Omega_\ast)$ and $C_{n}^{(m)}(\Omega_\ast)$ are given in Appendix~\ref{appendix-C}.
The effective pumping velocity ${\bm V}^{(3)} = - (\tau_{\rm p} / \overline{\rho}) \, \left\langle \tau \,{\bm u} \,  \Delta P \right\rangle$ is given by
\begin{eqnarray}
&& {\bm V}^{(3)}  = - {2\mu_0 \, D_T  \over 3(1 +\varepsilon_u)} \biggl\{2 A_{1}^{(-1)}(\Omega_\ast)  {\bm \nabla}
+ \biggl[A_{2}^{(-1)}(\Omega_\ast)
\nonumber\\
&& \; +  {3\varepsilon_u \over 4} \, \ln \, {\rm Re} \biggr]  {\bm \nabla}_\perp
- \Omega_\ast \, A_{3}^{(0)}(\Omega_\ast)  \, \hat {\bm \Omega} {\bm \times} {\bm \nabla}_\perp\biggr\}
\nonumber\\
&& \; \times \ln \left(\overline{T} \, \overline{P}^{\, 1/\gamma-1}\right) .
\label{BCLG6}
\end{eqnarray}
The effective pumping velocity  ${\bm V}^{(4)} = - (\tau_{\rm p} / \overline{\rho}) \, \langle \tau \,{\bm u} \,  (\hat{\bm \Omega} \cdot {\bm \nabla})^2 P \rangle$ is given by:
\begin{eqnarray}
&& {\bm V}^{(4)}  = - {2\mu_0 \, D_T  \over 15\, (1 +\varepsilon_u)} \biggl[\Big(5 A_{3}^{(-1)}(\Omega_\ast)
- C_{4}^{(0)}(\Omega_\ast)\Big) \, {\bm \nabla}
\nonumber\\
&& \; + \Big(A_{2}^{(0)}(\Omega_\ast) - 2 A_{1}^{(0)}(\Omega_\ast)
+ 12 C_{1}^{(0)}(\Omega_\ast)\Big) \, \hat {\bm \Omega} \Big(\hat {\bm \Omega} \cdot{\bm \nabla}\Big)
\nonumber\\
&& \; - {\Omega_\ast  \over 2} \Big(A_{2}^{(0)}(\Omega_\ast)
- A_{1}^{(0)}(\Omega_\ast) + 8 C_{1}^{(0)}(\Omega_\ast)\Big) \, \hat {\bm \Omega} {\bm \times} {\bm \nabla}_\perp\biggr]  .
\nonumber\\
&& \; \times \ln \left(\overline{T} \, \overline{P}^{\, 1/\gamma-1}\right) .
\label{CLG15}
\end{eqnarray}
The effective pumping velocity ${\bm V}^{(5)} = - \langle \tau \,{\bm u} \,  (\hat{\bm \Omega} \cdot {\rm rot} \,  {\bm u}) \rangle$
is given by:
\begin{eqnarray}
&& {\bm V}^{(5)}  = - {\Omega_\ast \over 3(1 +\varepsilon_u)} \, \biggl\{\left(\hat {\bm \Omega} {\bm \times} {\bm \lambda}_\perp\right) \, \biggl[\Omega_\ast^{-1} \biggl({9 \over 2} \varepsilon_u  + A_{3}^{(1)}(\Omega_\ast) \biggr)
\nonumber\\
&& \;
- 2 \Omega_\ast\biggl(A_{2}^{(3)}(2\Omega_\ast) + 4C_{1}^{(3)}(2\Omega_\ast)- C_{4}^{(3)}(2\Omega_\ast) \biggr) \biggr]
- 3 \varepsilon_u {\bm \lambda}_\perp
\nonumber\\
&& \;
+ {\bm \lambda} \biggl[{1 \over 6} - A_{1}^{(2)}(2\Omega_\ast) + {43 \over 5} \, C_{4}^{(2)}(\Omega_\ast) - 4 C_{4}^{(2)}(2\Omega_\ast)  \biggr]
\nonumber\\
&& \;
+ \hat {\bm \Omega} \, \Big(\hat {\bm \Omega} \cdot {\bm \lambda}\Big) \biggl[5 - \Omega_\ast^{-1}A_{1}^{(2)}(2\Omega_\ast) + {1 \over 2} \biggl(3 A_{3}^{(2)}(2\Omega_\ast)
\nonumber\\
&& \;
- 8 C_{1}^{(2)}(2\Omega_\ast) + 13 C_{4}^{(2)}(2\Omega_\ast) \biggr) - {54 \over 5} C_{4}^{(2)}(\Omega_\ast) \biggr]
+ \biggl[{1 \over 4}
\nonumber\\
&& \;
+ {37 \over 10} \, C_{4}^{(2)}(\Omega_\ast) - 2 C_{4}^{(2)}(2\Omega_\ast)\biggr] {\bm \nabla}
+ 3\biggl[1 + {1 \over 4} \biggl(A_{3}^{(2)}(2\Omega_\ast)
\nonumber\\
&& \;
+ 3C_{4}^{(2)}(2 \Omega_\ast) \biggr)
- {6 \over 5} \, C_{4}^{(2)}(\Omega_\ast) \biggr] \hat {\bm \Omega} \Big(\hat {\bm \Omega} \cdot {\bm \nabla}\Big)
\nonumber\\
&& \;+{1 \over 2} \,  \Omega_\ast^{-1}\biggl[3 - A_{3}^{(1)}(\Omega_\ast)  \biggr] \, \hat {\bm \Omega} {\bm \times} {\bm \nabla}_\perp\biggr\} D_T .
\label{CLG9}
\end{eqnarray}

For $\omega^2 \ll 1$, the total effective pumping velocity is given by:
\begin{eqnarray}
&& {\bm V}^{(\rm eff)} = (1 +2 \omega^2) \, {\bm V}^{(1)} + (1 - 2 \omega^2) \, {\bm V}^{(3)}
\nonumber\\
&& \; - \omega^2 \, \biggl[{\bm V}^{(2)} -3 {\bm V}^{(4)}
 + \hat {\bm \Omega} \, \Big[\hat {\bm \Omega} \cdot \Big({\bm V}^{(1)} + {\bm V}^{(3)} \Big) \biggr] .
 \nonumber\\
\label{CLS3}
\end{eqnarray}

\section{Effect of rotation on turbulence}
\label{appendix-B}

In this Appendix we derive equation for the
rotational contributions to the Reynolds stress and turbulent heat flux.
We apply the approach developed in \cite{RK18,RK19}.
Equations for fluctuations of velocity ${\bm u}'$ and entropy $s'$
are rewritten in the ${\bm k}$ space using new variables for
fluctuations of velocity ${\bm U}= \sqrt{\overline{\rho}}
\, {\bm u}'$ and entropy $s = \sqrt{\overline{\rho}} \,s'$. 
Using these equations we derive equations for the following
correlation functions:
$f_{ij}({\bm k},{\bm K}) = \langle U_i(t,{\bm k}_1) U_j(t,{\bm k}_2) \rangle$,
$F_{i}({\bm k},{\bm K}) = \langle s(t,{\bm k}_1) U_i(t,{\bm
k}_2) \rangle$, and
$\Theta_{i}({\bm k},{\bm K}) = \langle s(t,{\bm k}_1) s(t,{\bm
k}_2) \rangle$.
We apply the multi-scale approach \cite{RS75}, where
${\bm k}_1 = {\bm k} + {\bm K} / 2$, $\,{\bm k}_2 = -{\bm k} + {\bm K} / 2$, the wave vector ${\bm
K}$ and the vector ${\bm R}= ({\bm x}+{\bm y})/2$
correspond to the large-scale variables,  while ${\bm k}$ and ${\bm r}= {\bm y}-{\bm x}$
correspond to the  small-scale variables. Hereafter we omit
argument $t$ in the correlation functions. The
equations for these correlation functions are
given by
\begin{eqnarray}
{\partial f_{ij}({\bm k},{\bm K}) \over \partial t} =
L_{ijmn}^{\Omega} f_{mn} + M_{ij}^F + \hat{\cal N} \tilde f_{ij} ,
\label{CKB3}
\end{eqnarray}

\begin{eqnarray}
{\partial F_{i}({\bm k},{\bm K}) \over \partial t} &=&
D_{im}^{\Omega} F_{m} + g e_m P_{im}({\bm k}_1) \Theta
+ \hat{\cal N} \tilde F_{i} ,
\label{CKB4}
\end{eqnarray}

\begin{eqnarray}
{\partial \Theta({\bm k},{\bm K}) \over \partial t} &=&
\hat{\cal N} \Theta ,
\label{CKB5}
\end{eqnarray}
where $D_{ij}^{\Omega}({\bm k}) = 2 \varepsilon_{ijm} \Omega_n k_{mn}$,
$L_{ijmn}^{\Omega} = D_{im}^{\Omega}({\bm k}_1) \, \delta_{jn} +
D_{jn}^{\Omega}({\bm k}_2) \, \delta_{im}$,
$\delta_{ij}$ is the Kronecker unit tensor, $k_{ij} = k_i  k_j /
k^2$, $\varepsilon_{ijk}$ is the Levi-Civita fully antisymmetric tensor,
$P_{ij} ({\bm k})=\delta_{ij}-k_{ij}$,
and
\begin{eqnarray}
&& M_{ij}^F = g e_m [P_{im}({\bm k}_1) F_{j}({\bm
k},{\bm K})  + P_{jm}({\bm k}_2) F_{i}(-{\bm
k},{\bm K})],
\nonumber\\
\label{CEC3}
\end{eqnarray}

In Eqs.~(\ref{CKB3})--(\ref{CKB5}), $\hat{\cal N}\tilde f_{ij}$ and $\hat{\cal
N}\tilde F_{i}$ are the third-order moments
appearing due to the nonlinear terms $U^{N}_{m}$ and $s^{N}$, which are given by
\begin{eqnarray}
\hat{\cal N}\tilde f_{ij} &=& \langle P_{im}({\bm k}_1)
U^{N}_{m}({\bm k}_1) U_j({\bm k}_2) \rangle
\nonumber\\
&& + \langle U_i({\bm k}_1) P_{jm}({\bm k}_2) U^{N}_{m}({\bm k}_2)
\rangle ,
\label{CEC5}
\end{eqnarray}
\begin{eqnarray}
\hat{\cal N}\tilde F_{i} &=& \langle s^{N}({\bm k}_1) U_j({\bm k}_2)
\rangle + \langle s({\bm k}_1) P_{im}({\bm k}_2) U^{N}_{m}({\bm
k}_2) \rangle .
\nonumber\\
\label{CEC6}
\end{eqnarray}

\begin{eqnarray}
\hat{\cal N}\Theta = \langle s^{N}({\bm k}_1) s({\bm k}_2) \rangle
+ \langle s({\bm k}_1) s^{N}({\bm k}_2) \rangle ,
\label{CEC7}
\end{eqnarray}

The derived equations for the second-order moments of the velocity
fluctuations $\langle U_i \, U_j \rangle$, the
entropy fluctuations $\langle s^2 \rangle$ and
the turbulent flux of entropy $\langle s U_i
\rangle$,
include the first-order spatial differential operators
$\hat{\cal N}$  applied to the third-order moments $M^{(III)}$.
A problem arises how to close the system of the second-moment equations,
i.e., how to express the set of the third-moment terms
$\hat{\cal N} M^{(III)}({\bm k})$ through the lower moments
(see, e.g., \cite{O70,RI21}).
In the present study we use the spectral $\tau$ approximation
(see, e.g., \cite{O70,PFL76,KRR90,RI21})), which postulates
that the deviations of the third-moment terms, $\hat{\cal N} M^{(III)}({\bm k})$,
from the contributions to these terms afforded by the background rotating turbulence,
$\hat{\cal N} M^{(III,0)}({\bm k})$, are expressed through the similar
deviations of the second-order moments, $M^{(II)}({\bm k})-M^{(II,0)}({\bm k})$
in the relaxation form:
\begin{eqnarray}
&& \hat{\cal N}M^{(III)}({\bm k}) - \hat{\cal N}M^{(III,0)}({\bm k})
\nonumber\\
&& \quad \quad \quad \quad \quad
= -{M^{(II)}({\bm k}) - M^{(II,0)}({\bm k}) \over \tau_r(k)} .
\label{CCB6}
\end{eqnarray}
Here the correlation functions with the superscript $(0)$
correspond to the background non-rotating turbulence.
The time $\tau_r (k)$ is the characteristic relaxation time
of the statistical moments, which  can be identified with the
correlation time $\tau(k)$ of the turbulent
velocity field for large Reynolds numbers.

The rotational contributions to the Reynolds stress $f_{ij}^{(\Omega)} \equiv \langle U_i({\bm k}_1) \, U_j({\bm k}_2)  \rangle$
and turbulent heat flux $F_{i}^{(\Omega)} \equiv \langle s({\bm k}_1) \, U_i({\bm k}_2)  \rangle$
follows from Eqs.~(\ref{CKB3})--(\ref{CCB6}). We also take into account that the characteristic time of variations
of the second moments are much larger than the
correlation time $\tau(k)$ of the turbulent
velocity field.
Therefore, the rotational contributions to the Reynolds stress
and turbulent heat flux are given by  \cite{RK18,RK19,EGKR05}:
\begin{eqnarray}
f_{ij}^{(\Omega)} ({\bm k}) &=& L_{ijmn}^{-1} \, \Big[f^{(0)}_{mn} ({\bm k})  + \tau \, (L_{mnpq}^{\nabla}
+ L_{mnpq}^{\lambda}
\nonumber\\
&&+ \tilde M_{mn}^F) f^{(0)}_{pq} ({\bm k}) \Big] ,
\label{CLF1}
\end{eqnarray}
\begin{eqnarray}
F_{i}^{(\Omega)} ({\bm k}) = D_{ij}^{-1} \,  F^{(0)}_{j} ({\bm k}) ,
\label{CLFF1}
\end{eqnarray}
where
\begin{eqnarray}
&& D_{ij}^{-1} = \chi(\psi) \, (\delta_{ij} +
\psi \, \varepsilon_{ijm} \, \hat k_m + \psi^2 \, k_{ij}),
\label{CLF2}
\end{eqnarray}
\begin{eqnarray}
&& L_{ijmn}^{-1}({\bm \Omega}) =  {1 \over 2} \Big[B_1
\, \delta_{im} \delta_{jn} + B_2 \, k_{ijmn} +
B_3 \, (\varepsilon_{imp} \delta_{jn}
\nonumber\\
&& + \varepsilon_{jnp} \delta_{im}) \hat k_p +
B_4 \, (\delta_{im} k_{jn} + \delta_{jn} k_{im})
+ B_5 \, \varepsilon_{ipm} \varepsilon_{jqn}
k_{pq}
\nonumber\\
&& + B_6 \, (\varepsilon_{imp} k_{jpn} +
\varepsilon_{jnp} k_{ipm}) \Big] ,
\label{CLF3}
\end{eqnarray}

\begin{eqnarray}
\tilde M_{ij}^F &=& g e_m \Big[\big(P_{im}({\bm
k}) + k_{im}^\lambda \big)\tilde F_{j}({\bm k})+ \big(P_{jm}({\bm k})
\nonumber\\
&& - k_{jm}^\lambda\big) \tilde F_{i}(-{\bm k})\Big],
\label{CKBB8}
\end{eqnarray}
$\tilde F_{i}=F_{i}-F_{i}^{\Omega=0}$.
Here $\hat k_i = k_i / k$, $\, \chi(\psi) = 1 / (1
+ \psi^2) $, $\, \psi = 2 \tau(k) \, ({\bm k}
\cdot {\bm \Omega}) / k $, $\, B_1 = 1 + \chi(2
\psi) ,$ $\, B_2 = B_1 + 2 - 4 \chi(\psi) ,$ $\,
B_3 = 2 \psi \, \chi(2 \psi) ,$ $\, B_4 = 2
\chi(\psi) - B_1 ,$ $\, B_5 = 2 - B_1 $ and $B_6
= 2 \psi \, [\chi(\psi) - \chi(2 \psi)]$, see \cite{EGKR05}.
In Eqs.~(\ref{CLF1})--(\ref{CLFF1}), the operator $D_{ij}^{-1}$ is the inverse of
$\delta_{ij} - \tau \tilde D_{ij}$ and the
operator $L_{ijmn}^{-1}({\bm \Omega})$ is the
inverse of $\delta_{im} \delta_{jn} - \tau \,
\tilde L_{ijmn}$,
where $\tilde D_{ij} = 2 \varepsilon_{ijp} \Omega_q k_{pq}$ and
\begin{eqnarray}
\tilde L_{ijmn} &=& 2 \, \Omega_q \,
(\varepsilon_{imp}  \, \delta_{jn} +
\varepsilon_{jnp} \, \delta_{im}) \, k_{pq} .
\label{CLF37}
\end{eqnarray}
In Eq.~(\ref{CLF1}), the operators $L_{ijmn}^\nabla$ and $L_{ijmn}^\lambda$ are given by
\begin{eqnarray}
L_{ijmn}^\nabla = - 2\,\Omega_q \,
(\varepsilon_{imp}  \, \delta_{jn} -
\varepsilon_{jnp} \, \delta_{im}) \,
k_{pq}^\nabla ,
\label{CLF33}
\end{eqnarray}

\begin{eqnarray}
L_{ijmn}^\lambda &=& - 2\,\Omega_q \,
\Big[(\varepsilon_{imp}  \, \delta_{jn} -
\varepsilon_{jnp} \, \delta_{im}) \,
k_{pq}^\lambda
\nonumber\\
&&
+{2i \over k^2} (\varepsilon_{ilq} \,
\delta_{jn}   \,\lambda_m- \varepsilon_{jlq} \,
\delta_{im}\,\lambda_n) \, k_{l} \Big],
\label{CLF34}
\end{eqnarray}
where
\begin{eqnarray}
k_{ij}^\nabla &=& {i \over 2 k^2} \, [k_i \nabla_j
+ k_j \nabla_i - 2 k_{ij} ({\bm k} {\bm \cdot} \bec{\nabla})],
\label{CLF35}
\end{eqnarray}

\begin{eqnarray}
k_{ij}^\lambda = {i \over 2k^2} \, [k_i \lambda_j
+ k_j \lambda_i - 2 k_{ij} ({\bm k} {\bm \cdot} \bec{\lambda})] .
\label{CLF36}
\end{eqnarray}
In Eqs.~(\ref{CLF1})--(\ref{CLFF1}) we neglected effects ${\rm O}({\bm \lambda}^2, {\bm \Lambda}^2)$,
i.e., they are quadratic in fluid density stratification (described by ${\bm \lambda}$)
and inhomogeneity of turbulence (described by ${\bm \Lambda} = {\bm \nabla} \langle {\bm u}^2 \rangle/
\langle {\bm u}^2 \rangle$). We also take into account that $L_{ijmn}^{-1} \, P_{mn} ({\bm k}) = P_{ij} ({\bm k})$.

We use the following model of the background  inhomogeneous and density stratified
turbulence $f_{ij}^{(0)}({\bm k})  \equiv \langle U_i({\bm k}_1) \,
U_j({\bm k}_2)  \rangle^{(0)}$, which takes into account an
anisotropy of turbulence with appearance of rotation:
\begin{eqnarray}
&& f_{ij}^{(0)}({\bm k})  = {E(k)
\, [1 + 2 k \, \varepsilon_u \, \delta(\hat{\bm k}\cdot \hat {\bm \Omega}) \, \delta(\hat {\bm \Omega}\cdot \hat{\bm \nabla})]\over
8 \pi \, k^2 \, (1 + \varepsilon_u)}
\, \biggl[P_{ij}(k)
\nonumber\\
&& \quad + {{\rm i} \over k^2}\, \Big(\tilde\lambda_i \,
k_j - \tilde\lambda_j \, k_i\Big) \biggr] \langle {\bm U}^2 \rangle ,
\label{CLF4}
\end{eqnarray}
and the correlation function $F^{(0)}_{i}({\bm k}) = \langle s({\bm k}_1) \, U_i({\bm k}_2) \rangle^{(0)}$ is
\begin{eqnarray}
F^{(0)}_{i}({\bm k})  = - \tau(k) \tilde f_{ij}^{(0)}({\bm k}) \, \nabla_j \overline{S} ,
\label{CLG4}
\end{eqnarray}
where the function $\tilde f_{ij}^{(0)}({\bm k})$ is given by
Eq.~(\ref{CLF4}) without the terms proportional to $\tilde{\bm \lambda} = ({\bm \lambda}-{\bm \nabla})/2$.
Here we take into account that in the anelastic approximation the velocity
fluctuations ${\bm U}= \sqrt{\overline{\rho}}\, {\bm u}'$ satisfy the equation
$\bec{\nabla} \cdot {\bm U} = {\bm U} \cdot \bec{\bar\lambda}$, where
$\bec{\bar\lambda} \equiv \bec{\lambda} / 2= -(\bec{\nabla} \overline{\rho}) / 2\overline{\rho}$.

Equation~(\ref{CLF4}) is derived using symmetry arguments.
For instance, density stratification is taken into account in anelastic approximation
div $(\meanrho {\bm u}) = 0$ for low-Mach number turbulence.
In this case the derivation of the Reynolds stress in the Fourier space
$f_{ij}^{(0)} \equiv \langle U_i({\bm k}_1) \, U_j({\bm k}_2)  \rangle$ for a density-stratified turbulence
is described in Section~2.2.3 in Ref.~\cite{RI21}.
The derivation of the Reynolds stress in the Fourier space for an inhomogeneous turbulence
is described in Section~3.4.3 in Ref.~\cite{RI21}.
In particular, the large-scale effects caused by stratification ${\bm \lambda}$ and inhomogeneous turbulence ${\bm \nabla}\langle {\bm u}^2\rangle$, satisfy the conditions $\ell_0 \ll H_\rho$ and $\ell_0 \ll L_u$,
where $H_\rho$ and $L_u$ is the characteristic scales of the density stratification and inhomogeneity of turbulence, respectively.
Therefore, we consider only linear effects in ${\bm \lambda}$ and ${\bm \nabla}\langle {\bm u}^2\rangle$ in Eq.~(\ref{CLF4}),
so that the second rank tensor $f_{ij}^{(0)}$ is constructed as a linear combination of symmetric tensors, $\delta_{ij}$ and $k_{ij}$, with respect to the indexes $i$ and $j$, and non-symmetric tensors, $k_i \lambda_j$, $ k_j \lambda_i$, and $k_i \nabla_j\langle {\bm u}^2\rangle$, $ k_j \nabla_i\langle {\bm u}^2\rangle$.
To determine unknown coefficients multiplying by these tensors,
we use the following conditions in the derivation of Eq.~(\ref{CLF4}):
\begin{eqnarray}
\left\langle {\bm u}^2 \right\rangle = \int f_{ii}^{(0)} ({\bm k},{\bm K}) \, \exp({\rm i} \, {\bm K} {\bm \cdot} {\bm R}) \, d {\bm k}  \, d {\bm K} ,
 \label{RZ1}
\end{eqnarray}
\begin{eqnarray}
\left\langle\left({\rm div} \, {\bm u}\right)^2\right\rangle  = \int k_i \, k_j \, f_{ij}^{(0)} ({\bm k},{\bm K}) \, \exp({\rm i} \, {\bm K} {\bm \cdot} {\bm R}) \, d {\bm k} \, d {\bm K}.
\nonumber\\
 \label{RZ4}
\end{eqnarray}
The anisotropy of turbulence is caused by a fast rotation. The anisotropic turbulence can be described as a combination of a three-dimensional isotropic turbulence and two-dimensional turbulence in the plane perpendicular to the rotational axis. The degree of anisotropy $\varepsilon_u$ is defined as  the ratio of turbulent kinetic energies of two-dimensional to three-dimensional motions.
The parameter $\varepsilon_u$ is non-zero only for a fast rotation $\Omega_\ast \gg 1$.

In Eq.~(\ref{CLF4}), $\delta(x)$ is the Dirac delta function.
For a fast rotation, ${\bm \lambda}$ is perpendicular to ${\bm \Omega}$.
The turbulent correlation time in the ${\bm k}$ space
is $\tau(k) = 2\tau_0 \, \tilde \tau(k)$,
where $\tilde\tau(k) =(k / k_{0})^{1-q}$.
We assume that the background turbulence is of Kolmogorov type with
constant flux of energy over the spectrum,
i.e., the kinetic energy spectrum function for the
range of wave numbers $k_0<k<k_\nu$ is
$E(k) = - d \tilde \tau(k) / dk$, the function $\tilde\tau(k) =
(k / k_{0})^{1-q}$ with $1 < q < 3$ being the
exponent of the kinetic energy spectrum ($q =5/3$ for a Kolmogorov spectrum).
Here $k_{\nu} = 1 / \ell_{\nu}$
is the wave number based on the viscous scale $\ell_{\nu}$,
and $k_{0} = 1 / \ell_{0} \ll k_\nu$,
where $\ell_{0}$ is the integral (energy containing) scale of turbulent motions.

Note that a direct effect of rotation on turbulence is described self-consistently by Eqs.~(\ref{CKB3})--(\ref{CKB4}).
The kinetic helicity is produced by the natural way due to the combined effect
of uniform rotation and density stratification or inhomogeneity of turbulence.
The fast rotation is included in the background turbulence only
via the term proportional to $\varepsilon_u$ to take into account
appearance of strong anisotropy of background turbulence caused by the fast rotation.
This is "indirect" effect of transition from three-dimensional nearly
isotropic turbulence for slow rotation to a strongly anisotropic nearly
two-dimensional turbulence.

Using Eqs.~(\ref{CLF1})--(\ref{CLFF1}), we obtain $f_{ij}^{(\Omega,{\rm a})}({\bm k})
= L_{ijmn}^{-1} \, f^{(0)}_{mn} ({\bm k})$ and $F_{i}^{(\Omega)} ({\bm k}) = D_{ij}^{-1} \,  F^{(0)}_{j} ({\bm k}) $ as
\begin{eqnarray}
&& f_{ij}^{(\Omega,{\rm a})}({\bm k}) = {E(k)
\, [1 + 2 k \, \varepsilon_u \, \delta(\hat{\bm k}
\cdot \hat {\bm \Omega}) \, \delta(\hat {\bm \Omega}\cdot \hat{\bm \nabla})] \over
8 \pi \, k^2 \, (1 +\varepsilon_u)}
\, \biggl[P_{ij}(k)
\nonumber\\
&& + {{\rm i} \over k} \chi(\psi) \Big[\tilde\lambda_i \, \hat k_j- \tilde\lambda_j \, \hat k_i
+ \psi  \tilde\lambda_n \, \big(\varepsilon_{inp} \, k_{jp}
 - \varepsilon_{jnp} k_{ip}\big) \Big] \biggr] \langle {\bm U}^2 \rangle,
\nonumber\\
\label{CLG5}
\end{eqnarray}

\begin{eqnarray}
&& F_{i}^{(\Omega)}({\bm k}) = - {\tau(k) \, E(k)
\, [1 + 2 k \, \varepsilon_u \, \delta(\hat{\bm k}
\cdot \hat {\bm \Omega})] \over 8 \pi \, k^2 \, (1 +\varepsilon_u)} \, \left(\nabla_j \overline{S}\right)
\nonumber\\
&& \quad \times \chi(\psi) \, \biggl[P_{ij}(k) + \psi  \, \varepsilon_{ijs} \, \hat k_s \biggr]
\langle {\bm U}^2 \rangle  .
\label{CLG6}
\end{eqnarray}
Using Eqs.~(\ref{CLG6}) and~(\ref{CLG2})--(\ref{CLG3}), we determine
$\langle \tau u_i({\bm x})\, \big[\bec{\nabla}^2 \theta({\bm x})\big] \rangle$ as
\begin{eqnarray}
&& \left\langle \tau u_i({\bm x})\, \big[\bec{\nabla}^2 \theta({\bm x})\big] \right\rangle =
{1 \over 1 +\varepsilon_u} \, \biggl[2 A_{1}^{(-1)}(\Omega_\ast)  \, \delta_{ij}
\nonumber\\
&& \quad  + \biggl(A_{2}^{(-1)}(\Omega_\ast) +  {\varepsilon_u \over 2} \, \ln \, {\rm Re} \biggr) \Big(\delta_{ij} - \hat \Omega_{ij} \Big)
\nonumber\\
&& \quad + \hat \Omega_{m} \, \varepsilon_{ijm} \,
\Omega_\ast \, A_{3}^{(0)}(\Omega_\ast)  \biggr] \, \nabla_j  \ln \left(\overline{T} \, \overline{P}^{\, 1/\gamma-1}\right) .
\label{CLLLG5}
\end{eqnarray}
The functions $A_{n}^{(m)}(\Omega_\ast)$ are given in Appendix~\ref{appendix-C}.

\section{The identities used for the integration in the ${\bm k}$ space}
\label{appendix-C}

To integrate over the angles in the ${\bm k}$ space we used the
following identities:
\begin{eqnarray}
\bar I_{ij}(a) = \int {k_{ij} \sin \theta \over 1 + a \cos^{2}
\theta} \,d\theta \,d\varphi =  \bar A_{1}(a) \delta_{ij} + \bar
A_{2}(a) \, \Omega_{ij} ,
\nonumber\\
\label{APP1}
\end{eqnarray}

\begin{eqnarray}
\bar I_{ijmn}(a) &=& \int {k_{ijmn} \sin \theta \over 1 + a
\cos^{2} \theta} \,d\theta \,d\varphi = \bar C_{1} \Delta_{ijmn}
\nonumber\\
& &+ \bar C_{2} \, \Omega_{ijmn}
+ \bar C_{3} \, \Delta_{ijmn}^{(\Omega)} ,
\label{APP1C}
\end{eqnarray}
where $ \Omega_{ij} = \hat \Omega_{i} \hat \Omega_{j}$, and $\Omega_{ijmn}=\Omega_{ij} \, \Omega_{mn}$,
\begin{eqnarray}
\Delta_{ijmn} =\delta_{ij} \, \delta_{mn} + \delta_{im} \, \delta_{jn} + \delta_{in} \, \delta_{jm} ,
\label{APP2C}
\end{eqnarray}

\begin{eqnarray}
&& \Delta_{ijmn}^{(\Omega)} = \delta_{ij} \Omega_{mn}
+ \delta_{im} \Omega_{jn} + \delta_{in} \Omega_{jm} +
\delta_{jm} \Omega_{in}
\nonumber\\
& &
+ \delta_{jn} \Omega_{im} + \delta_{mn} \Omega_{ij} ,
\label{APP3C}
\end{eqnarray}
and the functions $\bar A_{k}(a)$ and $\bar C_{k}(a)$ are given by
\begin{eqnarray}
\bar A_{1}(a) = {2 \pi \over a} \biggl[(a + 1) \, {\arctan
(\sqrt{a}) \over \sqrt{a}} - 1 \biggr] ,
\label{APP2}
\end{eqnarray}

\begin{eqnarray}
\bar A_{2}(a) &=& - {2 \pi \over a} \biggl[(a + 3)  \,{\arctan
(\sqrt{a}) \over \sqrt{a}} - 3 \biggr]  .
\label{APP3}
\end{eqnarray}

\begin{eqnarray}
\bar C_{1}(a) &=& {\pi \over 2a^{2}} \biggl[(a + 1)^{2} {\arctan
(\sqrt{a}) \over \sqrt{a}} - {5 a \over 3} - 1 \biggr] ,
\label{APPN4}
\end{eqnarray}

\begin{eqnarray}
\bar C_{2}(a) &=& {\pi \over 2a^{2}} \biggl[(3 a^{2} + 30 a + 35)
{\arctan (\sqrt{a}) \over \sqrt{a}}
\nonumber\\
& & - {55 a \over 3} - 35 \biggr] ,
\label{APPN5}
\end{eqnarray}

\begin{eqnarray}
\bar C_{3}(a) &=& - {\pi \over 2a^{2}} \biggl[(a^{2} + 6 a + 5)
{\arctan (\sqrt{a}) \over \sqrt{a}} - {13 a \over 3} - 5 \biggr] .
\nonumber\\
\label{APPN6}
\end{eqnarray}
Note that $\bar A_{1} = 5 \bar C_{1} + \bar C_{3}$ and $\bar A_{2} =
\bar C_{2} + 7 \bar C_{3}$.

We introduce also the function $\bar A_{3}(a)=\bar A_{1}(a)+\bar A_{2}(a)$
and $\bar C_{4}(a)=\bar C_{1}(a)+\bar C_{3}(a)$:
\begin{eqnarray}
\bar A_{3}(a) &=& {4 \pi \over a} \biggl[1 -  {\arctan
(\sqrt{a}) \over \sqrt{a}} \biggr]  .
\label{APPN3}
\end{eqnarray}

\begin{eqnarray}
\bar C_{4}(a) &=& {2\pi \over a^{2}} \biggl[1 + {2 a \over 3}- (a+ 1)
{\arctan (\sqrt{a}) \over \sqrt{a}}\biggr] .
\nonumber\\
\label{APPN7}
\end{eqnarray}

These functions for $a \ll 1 $ are given by
\begin{eqnarray}
\bar A_{1}(a) = {4 \pi \over 3} \left(1 - {a \over 5} \right)  ,
\quad
\bar A_{2}(a) \sim - {8 \pi \over 15} \,  a ,
\label{APP4}
\end{eqnarray}
\begin{eqnarray}
\bar C_{1}(a) &\sim& {4 \pi \over 15} \left(1 - {a \over 7}\right)  ,
\quad
\bar C_{2}(a) \sim {32 \pi \over 315}  a^{2},
\label{APPP4}
\end{eqnarray}
\begin{eqnarray}
\bar C_{3}(a) &\sim& - {8 \pi \over 105} a ,
\quad
\bar C_{4}(a) \sim {4 \pi \over 15} \left(1 - {3a \over 7}\right) .
\label{APPP5}
\end{eqnarray}
These functions for $a \gg 1 $  are given by
\begin{eqnarray}
\bar A_{1}(a) &\sim& {\pi^{2} \over \sqrt{a}}  , \quad \bar A_{2}(a)
\sim - {\pi^{2} \over \sqrt{a}} ,
\label{APP5}
\end{eqnarray}
\begin{eqnarray}
\bar C_{1}(a) &\sim& {\pi^{2} \over 4 \sqrt{a}} - {4 \pi \over 3 a} ,
\quad
\bar C_{2}(a) \sim {3\pi^{2} \over 4 \sqrt{a}} ,
\label{APPP6}
\end{eqnarray}
\begin{eqnarray}
\bar C_{3}(a) &\sim& -{\pi^{2} \over 4 \sqrt{a}} + {8 \pi \over 3 a}  ,
\quad
\bar C_{4}(a) \sim {4 \pi \over 3 a} - {\pi^{2} \over a^{3/2}}   .
\label{APPP7}
\end{eqnarray}
Now we calculate the following functions
\begin{eqnarray}
A_{k}^{(p)}(\Omega_\ast) = {3 \over 4 \pi \, \Omega_\ast^{p+1}} \, \int_{0}^{\Omega_\ast}
Y^p \, \bar A_{k}(Y^{2}) \,d Y  ,
\label{APP6}
\end{eqnarray}

\begin{eqnarray}
C_{k}^{(p)}(\Omega_\ast) = {15 \over 4 \pi \, \Omega_\ast^{p+1}} \, \int_{0}^{\Omega_\ast}
Y^p \, \bar C_{k}(Y^{2}) \,d Y  ,
\label{APPN12}
\end{eqnarray}
where $p \geq 0$, and

\begin{eqnarray}
A_{k}^{(-1)}(\Omega_\ast) = {3 \over 4 \pi} \, \int_{\Omega_\ast {\rm Re}^{-1/2}}^{\Omega_\ast}
Y^{-1} \, \bar A_{k}(Y^{2}) \,d Y  ,
\label{APPM6}
\end{eqnarray}
where $\Omega_\ast=4 \Omega \, \tau_0$, $a=[\Omega_\ast \, \bar \tau(k)]^2=Y^2$ and $\bar \tau_\nu ={\rm Re}^{-1/2}$.
The integration yields:
\begin{eqnarray}
A_{1}^{(1)}(Z) &=& {3 \over 2} \biggl[{\arctan Z \over Z} \,
\biggl(1 - {1 \over Z^{2}} \biggr)
\nonumber\\
&& + {1 \over Z^{2}} \, \Big[1 - \ln(1 + Z^{2})\Big] \biggr] ,
\label{APP7}
\end{eqnarray}

\begin{eqnarray}
A_{2}^{(1)}(Z) &=& - {3 \over 2}  \biggl[{\arctan Z \over
Z} \biggl(1 - {3 \over Z^{2}} \biggr)
\nonumber\\
&& + {1 \over Z^{2}}\Big[3 - 2 \ln(1 + Z^{2})\Big] \biggr] ,
\label{APP8}
\end{eqnarray}

\begin{eqnarray}
A_{1}^{(2)}(Z) &=& {3 \over 4} \,\biggl[{\arctan Z \over Z} \,
\biggl(1 + {1 \over Z^{2}} \biggr) - {3 \over Z^{2}}
\nonumber\\
&& + {2 \over Z^{3}} S(Z) \biggr] ,
\label{APP9}
\end{eqnarray}

\begin{eqnarray}
A_{2}^{(2)}(Z) &=& - {3 \over 4} \, \biggl[{\arctan Z \over Z}
\biggl(1 + {1 \over Z^{2}} \biggr) - {7 \over Z^{2}}
\nonumber\\
&& + {6 \over Z^{3}} \, S(Z) \biggr],
\label{APP10}
\end{eqnarray}

\begin{eqnarray}
A_{3}^{(2)}(Z) &=& {3 \over Z^3} \, \biggl[Z - S(Z) \biggr],
\label{APPNN10}
\end{eqnarray}

\begin{eqnarray}
A_{1}^{(0)}(Z) &=& - {3 \over 4} \, \biggl[\arctan Z
\biggl(1 + {1 \over Z^{2}} \biggr) - {1 \over Z}
- 2 S(Z) \biggr],
\nonumber\\
\label{APP21}
\end{eqnarray}

\begin{eqnarray}
A_{2}^{(0)}(Z) &=& {9 \over 4} \, \biggl[\arctan Z \,
\biggl(1 + {1 \over Z^{2}} \biggr) - {1 \over Z}
- {2 \over 3} \, S(Z) \biggr],
\nonumber\\
\label{APP21}
\end{eqnarray}

\begin{eqnarray}
A_{3}^{(0)}(Z) &=& {3 \over 2} \, \biggl[\biggl(1 + {1 \over Z^{2}} \biggr)  \, \arctan Z  - {1 \over Z}\biggr] ,
\label{APP17}
\end{eqnarray}

\begin{eqnarray}
&& A_{1}^{(-1)}(Z) = {1 \over 2} \,\biggl[{8 \over 3} -{\arctan Z \over Z} \,
\biggl(3 + {1 \over Z^{2}} \biggr) + {1 \over Z^{2}}
\nonumber\\
&& \quad + \ln\left({{\rm Re} \over 1 + Z^{2}} \right) \biggr],
\nonumber\\
\label{APP14}
\end{eqnarray}

\begin{eqnarray}
A_{2}^{(-1)}(Z) &=& {3 \over 2} \, \biggl[{\arctan Z \over Z} \,
\biggl(1 + {1 \over Z^{2}} \biggr) - {1 \over Z^{2}} - {2 \over 3}\biggr],
\nonumber\\
\label{APP15}
\end{eqnarray}

\begin{eqnarray}
&& A_{3}^{(-1)}(Z) = {1 \over 3} \,\biggl[1 +{3 \over Z^{2}} \left({\arctan Z \over Z} -1\right)\biggr]
\nonumber\\
&& \quad +{1 \over 2} \, \ln\left({{\rm Re} \over 1 + Z^{2}} \right) ,
\label{APPNN14}
\end{eqnarray}

\begin{eqnarray}
C_{1}^{(2)}(Z) &=& {15 \over 16} \,\biggl[{\arctan Z  \over Z}\,
\biggl(1 - {1 \over Z^{4}} \biggr) - {13 \over 3Z^2}
\nonumber\\
&& + {1 \over Z^{4}} + {4 S(Z) \over Z^{3}} \biggr] ,
\label{APPNN19}
\end{eqnarray}

\begin{eqnarray}
C_{4}^{(2)}(Z) &=& {15 \over 4 Z^3} \,\biggl[\arctan Z  \,
\biggl(1 + {1 \over Z^{2}} \biggr) - {1 \over Z}
\nonumber\\
&& + {4 Z \over 3} - 2 S(Z) \biggr] ,
\label{APPNN9}
\end{eqnarray}

\begin{eqnarray}
C_{1}^{(3)}(Z) &=& {15 \over 8 Z^4} \,\biggl[{\arctan Z  \over Z}\,
\biggl({Z^{4} \over 3} + 2 Z^{2} - 1 \biggr) - Z^{2} + 1
\nonumber\\
&& - {4 \over 3} \ln \left(1 + Z^{2}\right) \biggr] ,
\label{BPPNN19}
\end{eqnarray}

\begin{eqnarray}
C_{1}^{(0)}(Z) &=& - {15 \over 32} \, \biggl[\arctan Z \,
\biggl(3 + {4 \over Z^{2}} + {1 \over Z^{4}}\biggr) - {11 \over 3Z}
\nonumber\\
&& - {1 \over Z^{3}} - 4 \, S(Z) \biggr],
\label{APPNN21}
\end{eqnarray}

\begin{eqnarray}
C_{4}^{(0)}(Z) &=& {15 \over 8} \,\biggl[\arctan Z  \,
\biggl(1 + {1 \over Z^{2}} \biggr)^2 - {5 \over 3Z} - {1 \over Z^3}\biggr] ,
\nonumber\\
\label{APPNN20}
\end{eqnarray}
where $ S(Z) = \int_{0}^{Z} [\arctan (Y) / Y] \,d Y$.
These functions for $ Z \ll 1 $ are given by
\begin{eqnarray*}
A_{1}^{(1)}(Z) &\sim& {1 \over 2} \left(1 - {Z^{2} \over 10} \right) ,
\;
A_{2}^{(1)}(Z) \sim - {Z^{2} \over 10} + {2 Z^{4} \over 35}  ,
\end{eqnarray*}
\begin{eqnarray*}
A_{3}^{(1)}(Z) &\sim& {1 \over 2} \left(1 - {3Z^{2} \over 10} \right) ,
\;
A_{3}^{(2)}(Z) \sim  {1 \over 3} \left(1 - {9Z^{2} \over 25} \right) ,
\end{eqnarray*}
\begin{eqnarray*}
A_{1}^{(2)}(Z) &\sim& {1 \over 3} \left(1 - {3Z^{2} \over 25} \right) ,
\;
A_{1}^{(0)}(Z) \sim Z \, \left(1 - {Z^{2} \over 15} \right) ,
\end{eqnarray*}
\begin{eqnarray*}
A_{2}^{(2)}(Z) &\sim& - {3Z^{2}\over 20} \left(1 - {4 Z^{2} \over 7}\right)  ,
\;
A_{2}^{(0)}(Z) \sim - {2 Z^3\over 15} ,
\end{eqnarray*}
\begin{eqnarray*}
A_{3}^{(0)}(Z) &\sim& Z \, \left(1 - {Z^{2} \over 5} \right) ,
\;
A_{2}^{(-1)}(Z)  \sim - {Z^{2} \over 5} ,
\end{eqnarray*}
\begin{eqnarray*}
A_{1}^{(-1)}(Z)  &\sim&  {1 \over 2} \,\left(\ln \, {\rm Re}  - {Z^{2} \over 5}\right)  ,
\end{eqnarray*}
\begin{eqnarray*}
A_{3}^{(-1)}(Z) &\sim&  {1 \over 2} \,\left(\ln \, {\rm Re}  + {Z^{2} \over 5}\right)  ,
\end{eqnarray*}
\begin{eqnarray*}
C_{1}^{(2)}(Z) &\sim& {1 \over 3} \left(1 - {3Z^{2} \over 35} \right) ,
\;
C_{4}^{(2)}(Z) \sim {1 \over 3} \left(1 - {9Z^{2} \over 35} \right) ,
\end{eqnarray*}
\begin{eqnarray*}
C_{1}^{(3)}(Z) &\sim& {1 \over 4} \left(1 - {2Z^{2} \over 21} \right) ,
\;
C_{1}^{(0)}(Z) \sim Z ,
\end{eqnarray*}
\begin{eqnarray*}
C_{4}^{(0)}(Z) &\sim& Z\,  \left(1 - {Z^{2} \over 7} \right) .
\end{eqnarray*}
These functions for $ Z \gg 1 $ are given by
\begin{eqnarray*}
A_{1}^{(1)}(Z) &\sim& {3 \pi \over 4Z} - {3 \ln Z \over Z^2} ,
\;
A_{2}^{(1)}(Z) \sim - {3 \pi \over 4Z} + {6 \ln Z \over Z^2} ,
\end{eqnarray*}
\begin{eqnarray*}
A_{3}^{(1)}(Z)  &\sim& {3 \ln Z \over Z^2},
\;
A_{3}^{(2)}(Z)  \sim {3 \over Z^2},
\end{eqnarray*}
\begin{eqnarray*}
A_{1}^{(2)}(Z) &\sim& {3\pi \over 8Z}\left(1 - {6 \over \pi Z} \right)  ,
\;
A_{2}^{(2)}(Z) \sim -{3\pi \over 8Z}\left(1 - {14 \over \pi Z} \right)  ,
\end{eqnarray*}
\begin{eqnarray*}
A_{1}^{(0)}(Z)  &\sim& {3\pi \over 4} \,  \left(\ln Z - {1 \over 2}\right),
\;
A_{2}^{(0)}(Z)  \sim - {3\pi \over 4} \,  \ln Z ,
\end{eqnarray*}
\begin{eqnarray*}
A_{3}^{(0)}(Z)  &\sim& {3 \over 4} \,  \left(\pi  - {2 \over Z} \right),
\;
A_{2}^{(-1)}(Z) \sim {3 \over 4} \,  \left({\pi \over Z} - {4 \over 3} \right) ,
\end{eqnarray*}
\begin{eqnarray*}
C_{1}^{(2)}(Z) &\sim& {15\pi \over 32Z} , \; C_{1}^{(3)}(Z) \sim {5\pi \over 16Z} ,
\;
C_{4}^{(2)}(Z) \sim {5 \over Z^2} ,
\end{eqnarray*}
\begin{eqnarray*}
C_{1}^{(0)}(Z) &\sim& {15 \pi \over 16} \,  \biggl(\ln Z - {3 \over 4}\biggr),
\; C_{4}^{(0)}(Z) \sim - {5 \over Z} + {15\pi \over 16}.
\end{eqnarray*}
and when $1 \ll Z^2 \ll {\rm Re}$, the function
\begin{eqnarray*}
A_{1}^{(-1)}(Z) \sim {1 \over 2} \,\left[\ln\left({{\rm Re} \over Z^{2}}\right)  + {8 \over 3}\right] ,
\\
A_{3}^{(-1)}(Z) \sim {1 \over 2} \,\left[\ln\left({{\rm Re} \over Z^{2}}\right)  + {2 \over 3}\right] .
\end{eqnarray*}

To integrate over the angles in ${\bm k}$-space for anisotropic part of turbulence, we use
the following integrals:
\begin{eqnarray}
&& \int k_{ij}^{\perp} \,d\varphi = \pi \delta_{ij}^{(2)},
\quad \int k_{ijmn}^{\perp} \,d\varphi = {\pi \over 4} \Delta_{ijmn}^{(2)},
\label{CLEC14}
\end{eqnarray}
where $\delta_{ij}^{(2)}\equiv P_{ij}(\Omega) = \delta_{ij} - \Omega_{ij}$ and
$\Delta_{ijmn}^{(2)} = P_{ij}(\Omega) P_{mn}(\Omega) + P_{im}(\Omega) P_{jn}(\Omega)
+ P_{in}(\Omega) P_{jm}(\Omega)$.

\end{document}